\documentclass[twocolumn]{aastex631}
\usepackage{amsmath}

\begin{document}

%\title{Formation of WNL stars for the MW and LMC based on the $k-\omega$ model}

\title{Formation of WNL stars for the MW and LMC based on the $k-\omega$ model}

\author[0009-0000-3338-0853]{Jijuan Si}
\affiliation{Yunnan Observatories, Chinese Academy of Sciences, P.O.Box 110, Kunming 650216, China}
\affiliation{University of Chinese Academy of Sciences, Beijing 100049, China}
\affiliation{Key Laboratory for Structure and Evolution of Celestial Objects, Chinese Academy of Sciences, People’s Republic of China}
\affiliation{International Centre of Supernovae, Yunnan Key Laboratory, Kunming 650216, China}

\author[0009-0000-3338-0853]{Zhi Li}
\affiliation{Yunnan Observatories, Chinese Academy of Sciences, P.O.Box 110, Kunming 650216, China}
\affiliation{University of Chinese Academy of Sciences, Beijing 100049, China}
\affiliation{Key Laboratory for Structure and Evolution of Celestial Objects, Chinese Academy of Sciences, People’s Republic of China}
\affiliation{International Centre of Supernovae, Yunnan Key Laboratory, Kunming 650216, China}

\author[0000-0002-1424-3164]{Yan Li}
\affiliation{Yunnan Observatories, Chinese Academy of Sciences, P.O.Box 110, Kunming 650216, China}
\affiliation{University of Chinese Academy of Sciences, Beijing 100049, China}
\affiliation{Key Laboratory for Structure and Evolution of Celestial Objects, Chinese Academy of Sciences, People’s Republic of China}
\affiliation{International Centre of Supernovae, Yunnan Key Laboratory, Kunming 650216, China}
\affiliation{Center for Astronomical Mega-Science, Chinese Academy of Sciences, Beijing 100012, China}



\begin{abstract}

We adopt a set of second-order differential equations ($k-\omega$ model) to handle core convective overshooting in massive stars, simulate the evolution of WNL stars with different metallicities and initial masses, both rotating and non-rotating models, and compare the results with the classical overshooting model. The results indicate that under the same initial conditions, the $k-\omega$ model generally produces larger convective cores and wider overshooting regions, thereby increasing the mass ranges and extending the lifetimes of WNL stars, as well as the likelihood of forming WNL stars. The masses and lifetimes of WNL stars both increase with higher metallicities and initial masses. Under higher-metallicity conditions, the two overshooting schemes significantly differ in their impacts on lifetimes of the WNL stars, but insignificant in the mass ranges of the WNL stars. Rotation may drive the formation of WNL stars in low-mass, metal-poor counterparts, with this effect being more pronounced in the OV model. The surface nitrogen of metal-rich WNL stars formed during the MS phase is likely primarily from the CN-cycle, while it may come from both the CN- and NO-cycles for relatively metal-poor counterparts. Our model can effectively explain the distribution of WNL stars in the Milky Way, but appears to have inadequacies in explaining the WNL stars in the LMC.

\end{abstract}

\keywords{WNL stars --- Metallicity  --- Convective overshooting --- Mass loss of the massive stars}

\section{Introduction} \label{sec:intro}
Massive stars are a unique type of stars that have drawn significant attention from astronomers. Renowned for their robust stellar winds and enigmatic convection phenomena, these stars wield a profound influence on their own evolutionary trajectories. The culmination of their life cycle often involves dramatic supernova explosions, giving rise to extraordinary high-energy phenomena such as black holes, neutron stars, pulsars, and magnetic stars \citep{2014ARA&A..52..487S}.
The core-collapse explosions of massive stars play a pivotal role in the production of heavy elements in the universe, serving as a key contributor to the enrichment of metals and dust within galaxies \citep{2015MNRAS.452.1068C}. The mass-loss phenomenon expels heavy elements in the form of gas or dust into the interstellar medium (ISM), serving as raw materials for the formation of the next generation of stars. 
However, the mass loss of massive stars due to line-driven processes remains unclear \citep{2023A&A...673A.109G}, constituting a significant uncertainty in the calculation of models for the massive stars \citep{2014ARA&A..52..487S}.

A distinctive subgroup among the massive stars is Wolf-Rayet (WR) star, first discovered and named by the French astronomers \citep{1867CRAS...65..292W}. The WR stars are accompanied by strong and broad emission lines \citep{2007ARA&A..45..177C}. According to \citet{201962557Hamann}, they tend to exhibit higher mass-loss rates ($\dot{M}$) than O-type stars. The mass loss for H-rich WN (nitrogen sequence WR) stars can reach values as high as $4 \times 10^{-5}$ $\mathrm{M}_{\odot} \mathrm{yr}^{-1}$, signifying the presence of powerful stellar winds. There is evidence that the winds of WR stars are highly heterogeneous \citep{1988ApJ...334.1038M, 2007ApJ...656L..77M}.
 
The typical mass of WR star ranges between $10-25$ $\mathrm{M}_{\odot}$, but some can reach over $300$ $\mathrm{M}_{\odot}$ (e.g., R136a1 is a WN5h star) \citep{2010MNRAS.408..731C}. This suggests that the initial masses of their predecessors are notably higher. 
According to the studies by \citet{1991A&A...241...77S}, \citet{2003A&A...404..975M}, and \citet{2012A&A...542A..29G}, the WR stars are characterized as stars with $\rm{log}$ ${(T_{\rm {eff}}/K)}>4.0$ and a surface H mass fraction $X_{\rm s} < 0.3$.
Classified based on their spectral characteristics, the WR stars are further divided into nitrogen (N) sequence Wolf-Rayet (WN), carbon (C) sequence Wolf-Rayet (WC), and oxygen (O) sequence Wolf-Rayet (WO) types. The subclass WN denotes a WR star with $X_{\rm s} > 10^{-5}$, corresponding to a late-type WN star (WNL). An early-type  WN star (WNE) is a WR star without H and with a surface C abundance lower than the N abundance (${\text C}_{\text s} < {\text N}_{\text s}$). Since several WNE samples with $X_{\rm s} > 10^{-5}$, we adopt the definition based on \citet{2006A&A...457.1015H}, categorizing the WN stars with $0.05 \le X_{\rm s} \le 0.4$ and $\rm{log}$ ${(T_{\text{eff}}/K)}>4.0$ as WNL stars, and those with lower hydrogen content $X_{\rm s} < 0.05$ as WNE stars. The other two WR subtypes are WC and WO stars, both identified by strong lines of He, C, and O, and characterized by a lack of H.

Normally, the WNL stars can either be classical (post main-sequence, hydrogen-rich) or very massive main-sequence WR stars \citep{2023A&A...674A..88D}. The former are usually known as core He-burning WR stars (classical WR or cWR, \citep{2019A&A...627A.151S}). The latter are less evolved very massive stars (VMS) and normally show WR-type spectra of the WNh subclass \citep{sander_vink_higgins_shenar_hamann_todt_2020}. These VMS typically have high luminosity and presence of H, and evolve into the WNL stage at the very start (i.e. the ZAMS). There is also evidence that the WNL stars may have originated from the most massive O-type stars \citep{2010MNRAS.408..731C} or even been born as the WNL stars \citep{2008A&A...478..219M, 2009A&A...495..257M, 2011A&A...535A..56G}.
As more bright WR stars are discovered, with 676 WR data from the Milky Way (MW) released in Gaia DR3 until May 2024 (Galactic Wolf-Rayet Catalogue) \footnote{\url{https://pacrowther.staff.shef.ac.uk/WRcat/index.php}}.

For the surface abundances and masses of WNL stars reflect the result of internal convection and external mass loss. It serves as a valuable platform for understanding the MS even He-MS evolutionary features of massive stars. Scholars have recently developed models to address the structure and evolution of massive stars. 

Convection and convective overshooting are key factors influencing the evolution and internal structure of the massive stars. By transporting a variety of chemical compounds into the burning zone and carrying nucleosynthesis products outward, it supports nuclear reactions and promotes the mixing of reaction products, thereby altering the chemical composition of the stellar surface. This changes the stellar lifetimes, representing one of the most significant uncertainties in the study of massive stars \citep{RevModPhys.74.1015, doi:10.1146/annurev-astro-081811-125534, 2019ApJ...870...77L}.

The standard mixing-length theory (MLT) \citep{1958ZA.....46..108B} is a common model for stellar convection. Classical diffusive-overshoot model is mostly based on an exponentially decaying law \citep{2000A&A...360..952H} in the overshooting region. Scholars have found that various values of the overshooting parameter also affect the width of the main sequence, according to \citet{2019ApJ...870...77L} and \citet{2023ApJS..268...51L}. Recently, \citet{2012ApJ...756...37L, 2017ApJ...841...10L} develops the $k-\omega$ model to describe the convection and overshooting in stars, incorporating the effects of turbulent kinetic energy diffusion. This model accurately captures the behavior of convective motions in both the core and envelope of the star. It has been demonstrated to be suitable for 30 $\mathrm{M}_{\odot}$ stellar models \citep{2019ApJ...870...77L} and low-mass stars \citep{Guo_2019}. Besides, the $k-\omega$ model has been used by \citet{Guo_2019} to calibrate the value of the free parameter $f_{\mathrm{ov}}$ for the low-mass stars. We compare the two models to investigate the overshooting results of WNL stars treated with different methods in this work.

Meanwhile, considering the impact of metallicity on mass loss from stellar winds is crucial. The strong correlation between the mass loss rates of massive stars and their chemical composition has been studied extensively by many scholars \citep{2000A&A...360..227N, 2005A&A...442..587V, 2008A&A...482..945G}, who have provided targeted prescriptions for various evolutionary stages. For O- and B-type stars, the prescription of mass-loss rate from \citet{2000A&A...362..295V, 2001A&A...369..574V} is commonly employed. The empirical mass-loss prescription of WR stars are studied by \citet{2000A&A...360..227N} and \citet{2008A&A...482..945G}. Additionally, mass loss also play an effect on the angular momentum transfer of the rotating stars. 

Furthermore, relevant results from studies on rotation models can be found in \citet{2003A&A...404..975M}, \citet{2012A&A...537A.146E}, \citet{2012A&A...542A..29G}, \citet{2016ApJ...823..102C}, and \citet{2019A&A...627A..24G}. The conclusion drawn from these studies suggests that rotation has a substantial impact on the transfer of angular momentum and mixing of stellar chemical components, as indicated by \citet{2012A&A...537A.146E}. In the Equation \ref{eq:Lwind}, where $\Omega_{\mathrm{s}}$ is the angular velocity of the surface, and $r_{*}$ the stellar radius, and $\Delta t$ is the current time step. It is obvious that the angular momentum loss due to the rotating stars are significant amount. Furthermore, \citet{2019A&A...627A..24G} presents that mixing and mass loss have the most effects on surface abundances. Rotation is an issue that cannot be disregarded since it will interact with convection to drastically alter the constituent abundance profile during the mixing process. All of these factors are important in the formation of the WNL stars.

\begin{equation}
\Delta \mathcal{L}_{\text {winds}}=\frac{2}{3} \dot{M} \Omega_{\mathrm{s}} r_{*}^{2} \Delta t.
\label{eq:Lwind} 
\end{equation}

While previous studies on the massive single stars have provided valuable insights, their conclusions are limited in generalizability due to variations in employed models and inputs. The previous model of stellar wind mass loss lacks sufficient correlation with metals, as well as the study of stellar internal overshoot is incomplete.

In present work, we plan to use the MESA (Module for Experiments in Stellar Astrophysics) software to simulate the evolutionary grid of the WNL single stars. Our simulations will incorporate key factors such as convective overshooting, rotation, and metallicity-dependent mass loss in the stellar wind. We aim to discuss the impact of the new prescription for dealing with the overshooting and mass loss of WNL stars, understand the influence of convective overshooting mixing in massive stars on their evolution, and provide a comprehensive description of the evolutionary characteristics of WNL stars, thereby expanding upon existing grid frameworks.
 
The structure of this paper is organized as follows:
Section~\ref{sec:method and input} provides an overview of the models and input parameters used in this study. 
Section~\ref{sec:output} presents an analysis of the output results.
Section~\ref{sec:observation} compares the results of model with the observations and discusses potential discrepancies between the models and observations.
Section~\ref{sec:sum} summaries the results modeled by the two overshooting schemes.
The Appendix shows the primary output of all initial parameter combinations when rotating.

\section{Method and input} \label{sec:method and input}
 
MESA is a suite of open-source libraries for various computational stellar astrophysics applications \citep{2010ascl.soft10083P}, developed by \citet{2011ApJS..192....3P, 2013ApJS..208....4P, 2015ApJS..220...15P, 2018ApJS..234...34P, 2019ApJS..243...10P}. In this study, we utilize the ``black\_hole'' package from MESA\footnote{\url{https://docs.mesastar.org/}} Version 12115 to simulate the evolution of massive single star until core He exhaustion. The simulation covers initial stellar masses ($M_{\mathrm i}$) ranging from $50$ $\mathrm{M}_{\odot}$ to $150$ $\mathrm{M}_{\odot}$ in steps of $10$ $\mathrm{M}_{\odot}$ at various metallicities ($Z = 0.002, 0.008, 0.014, 0.02, $ and $ 0.04$). We consider both cases with and without rotation.
 
\subsection{Convection and convective overshooting} \label{subsec:model}
The standard MLT is commonly used to deal with the region of stellar convection, but it does not account for turbulence properties such as diffusion and anisotropy. For the overshooting region outside the Schwarzschild boundary, exponentially decaying overshooting is a common approach, proposed by \citep{1997A&A...324L..81H, 2000A&A...360..952H}. However, it is the result of the direct numerical simulations. As concluded in \citet{2023ApJS..268...51L}'s work, the adjustable value of the free parameter $f_{\mathrm{ov}}$ depends on the initial mass at solar metallicity, which poses inconvenience and inaccuracy in describing the overshooting region for stellar models with different masses.

While the $k-\omega$ model proposed by \citep{2012ApJ...756...37L}, which is distinguished from the previous direct numerical simulations. This model is based on moment equations of fluid hydrodynamics \citep{2012ApJ...756...37L, 2017ApJ...841...10L, 2019ApJ...870...77L}, which is a second-order partial differential equation for the kinetic energy of turbulence and the frequency of turbulence. The assumptions of this stellar convection model about the convective vortex cells are very similar to the mean fluid element in the mixing length theory, which better describes the scale of the stellar convective cells \citep{2012ApJ...756...37L}. It allows for a set of parameters directly applicable to account for convection and convective overshooting across stars of varying masses. Furthermore, the improvement of this model not only can describe the stellar envelopes but also the stellar cores \citep{2017ApJ...841...10L}.

In the present paper, we adopt the Schwarzschild criterion to determine the boundary of the convection zone and primarily investigate the core convection and overshooting.

\subsubsection{Overshooting model} \label{subsubsec:OV model}
In order to compare the classical model with the $k-\omega$ model in dealing with convection and overshooting, we treat the convective zone with MLT and set the mixing-length parameter as $\alpha = 1.5$ according to \citet{1965ApJ...142..841H}. And the overshoot mixing beyond the convective zone is considered using a diffusion approach with an exponentially decaying diffusion coefficient based on the \citet{2000A&A...360..952H} approach (referred to as the OV model). The overshoot mixing diffusion coefficient, $D_{ov}$, decreases exponentially as a function of distance:
 
\begin{equation}
D_{\mathrm{ov}} = D_{\mathrm{conv},0} {\rm exp} (-\frac{2z}{f_{\mathrm{ov}}H_{\mathit P, 0}} ),
\label{eq:fov}
\end{equation}
Here, $D_{\mathrm conv,0}$ represents the diffusion coefficient at the convective core boundary, $z$ is the distance to the convective core boundary, $H_{\mathit P, 0}$ denotes the local pressure scale height, and $f_{\mathrm{ov}}$ is a free parameter governing the width of convective overshoot. The free parameter $f_{\mathrm{ov}}$ is set to 0.016 referred to \citet{2016ApJ...823..102C}.

\subsubsection{k-omega model} \label{subsubsec:KO model}  

We incorporate the $k-\omega$ model (referred to
as KO model) into the MESA ``run\_star\_extras.f'' package for managing core convection and overshooting. The $k-\omega$ model is primarily defined by two equations describing turbulent kinetic energy ($k$) and turbulence frequency ($\omega$) in Equation \ref{eq:k} and \ref{eq:omega}. 
\begin{equation}
\frac{\partial k}{\partial t}-\frac{1}{r^{2}} \frac{\partial}{\partial r}\left(r^{2} \nu_{t} \frac{\partial k}{\partial r}\right)=S+G-k \omega,
\label{eq:k}
\end{equation}

\begin{equation}
\frac{\partial \omega}{\partial t}-\frac{1}{r^{2}} \frac{\partial}{\partial r}\left(r^{2} \frac{\nu_{t}}{\sigma_{\omega}} \frac{\partial \omega}{\partial r}\right)=\frac{c_{L}^{2} k}{L^{2}}-\omega^{2},
\label{eq:omega}
\end{equation}
where $S$ and $G$ denote the shear and buoyancy production rate of the turbulent kinetic energy, respectively. 
The turbulent diffusivity ${\nu_{t}}$ is defined as $c_{\mathit \mu}$$\frac{k}{\omega}$. Based on \citet{2000tufl.book.....P}, the parameters of $c_{\mathit \mu}$, $c_{L}$, and $\sigma_{\mathit \omega}$ are assigned values equal to 0.09, $c_{\mathit \mu}^{3/4}$, and 1.5, as shown in Table~\ref{tab:para}. In the table, the parameter $c_{\mathit h}$ directly influences the turbulent heat flux, while $c_{\mathit X}$ controls the efficiency of overshoot mixing \citep{2017ApJ...841...10L}. 
$L$ represents the macro-length of turbulence, which is equivalent to the mixing length in the standard MLT.
If the thickness of the stellar convection zone is much greater than the local pressure scale height, the macro-length of turbulence $L$ usually assumed to be proportional to the local pressure scale height $H_{\mathit P}$ \citep{ 2017ApJ...841...10L, 2019ApJ...870...77L}. The $H_{\mathit P}$ is defined in \ref{eq:Hp}. While in the $k-\omega$ model, the treatment of the macro-length of turbulence in the core differs from that in the envelope. Because the size of the convective core is usually smaller than the local pressure scale height. Therefore, the macro-length of turbulence 
$L$ in the core is constrained by the actual size of the convection core: 

\begin{equation}
L = c_{L} \alpha^{'} R_{\mathrm {cc}}, 
\label{eq:L} 
\end{equation}
where $R_{\mathrm {cc}}$ is the radius of the convective core, the $\alpha^{'}$ is an adjustable parameter. The improvement of the the macro-length of turbulence L also distinguishes from the traditional MLT.

According to \citet{2012ApJ...756...37L, 2017ApJ...841...10L} and \citet{2019ApJ...870...77L}, the buoyancy production rate is approximated by: 
\begin{equation}
G=-\frac{c_{t}}{1+\frac{\lambda \omega}{\rho c_{P} k}+c_{t} c_{\theta} \omega^{-2} N^{2}} \frac{k}{\omega} N^{2},
\label{eq:G} 
\end{equation} 
The two parameters $c_{t}$ and $c_{\theta}$ are given by turbulence models \citep{2019ApJ...870...77L}: $c_{t}$ = 0.1 and $c_{\theta}$ = 0.5. The $c_{\mathit P}$ represents the specific heat at constant pressure. The variables $\rho$ and $P$ denote density and total pressure, respectively. In the Equation \ref{eq:G}, $\lambda$ is radiation diffusivity and is defined as: 
\begin{equation}
\lambda=\frac{16 \sigma T^{3}}{3 \rho \kappa},
\label{eq:lambda}
\end{equation}
where $\sigma$ is the Stefan– Boltzmann constant, $\kappa$ is Rosseland mean opacity.
 
\begin{deluxetable*}{ccccccccccccc} 
\tablecaption{Parameters of the $k-\omega$ model. \label{tab:para}}
\tablewidth{0pt}
\setlength{\tabcolsep}{6pt}
\tablehead{
         \colhead{$c_{\mathit \mu}$} &  \colhead{$\sigma_{\mathit \omega}$} &  \colhead{$c_{\mathit \theta}$} &   \colhead{$c_{\mathit t}$} &   \colhead{$c_{\mathit h}$} &   \colhead{$c_{\mathit X}$}&     \colhead{$\alpha^{'} $}
         }
%\decimalcolnumbers
\startdata
	0.09 &  1.5&  0.5&  0.1&  2.344&  1.0&    0.06 \\	
\enddata
\end{deluxetable*}

In the $k-\omega$ model, \citet{2012ApJ...756...37L, 2017ApJ...841...10L} defines the turbulent diffusivity $D_{\mathit t}$ as:
\begin{equation}
D_{\mathit t}=\frac{c_{X}}{1+\frac{\lambda \omega}{\rho c_{\mathit P} k}+c_{t} c_{\theta} \omega^{-2} N^{2}} \frac{k}{\omega},
\label{eq:D}
\end{equation}
The variation of $D_{t}$ is represents the diffusion coefficient due to the convective and overshoot mixing. We use the same values of $c_{t}$ and $c_{\theta}$ as \citet{2023ApJS..268...51L} and list in Table~\ref{tab:para}. The N means buoyancy frequency:
\begin{equation}
N^{2}=-\frac{\beta g T}{H_{\mathit P}}\left(\nabla-\nabla_{\mathrm{ad}}\right),
\end{equation}
 
\begin{equation}
H_{\mathit P} = -\frac{\mathrm{d} r}{\mathrm{d}\ln{\mathit P}}.
\label{eq:Hp}
\end{equation}
where ${\beta}$ is a thermodynamic coefficient. According to the G, the property of the stratification can be determined: if $G > 0$ in a convection zone and the $N^{2} < 0$, if $G < 0$ in a stably stratified region and the $N^{2} > 0$.

The evolution of the abundance of element ``i'' is given by diffusion equation:
\begin{equation}
\frac{\partial X_{i}}{\partial t}=\frac{\partial}{\partial m}\left[\left(4 \pi \rho r^{2}\right)^{2}\left(D_{\text {mix }}+D_{\mathit t}\right) \frac{\partial X_{i}}{\partial m}\right]+d_{i},
\end{equation}
where $X_{i}$ is the mass fraction of element ``i'', and $d_{i}$ is the generation rate of the element ``i''. $D_{\text {mix}}$ represents the diffusion coefficient contributed by various mixing processes, including element diffusion, meridional circulation, convective overshoot mixing, etc. For more information, please refer to \citep{2012ApJ...756...37L, 2017ApJ...841...10L}.

\subsection{Mass loss from stellar wind} \label{subsec:mass loss}
 
In this study, we specifically employ the ``Dutch'' mass loss scheme, which combines different prescriptions for hot and cool stars to handle various temperature regions in the HR diagram. 

We adopt the mass-loss prescription provided by \citet{2001A&A...369..574V} for stars with $\rm{log}$ $(T_{\text{eff}}/K) > 4.0$ and $X_{\rm s} > 0.4$. This prescription accounts for the mass loss on both sides of the bi-stability jumps of massive stars at specific temperatures, as shown in Equations\ref{eq:Mdot_Vink_hot} and \ref{eq:Mdot_Vink_cool} for the hot temperature domain and cool temperature domain, respectively.

For $27500 < T_{\text {eff}} \le 50000$ K,
\begin{align}
\label{eq:Mdot_Vink_hot}
\log \dot{M} = & -6.697( \pm 0.061) \notag \\
 & +2.194( \pm 0.021) \log \left(L_{*} / 10^{5}\right) \notag \\
 & -1.313( \pm 0.046) \log \left(M_{*} / 30\right) \notag \\
 & -1.226( \pm 0.037) \log \left(\frac{v_{\infty} / v_{\text {esc }}}{2.0}\right) \notag \\
 & +0.933( \pm 0.064) \log \left(T_{\text {eff }} / 40000\right) \notag \\
 & -10.92( \pm 0.90)\left\{\log \left(T_{\text {eff }} / 40000\right)\right\}^{2} \notag \\
 & +0.85( \pm 0.10) \log \left(Z / Z_{\odot}\right),
\end{align}

for $12500 \le T_{\text {eff}} \le 22500$ K,
\begin{align}
\label{eq:Mdot_Vink_cool}
\log \dot{M} = & -6.688( \pm 0.080) \notag \\
 & +2.210( \pm 0.031) \log \left(L_{*} / 10^{5}\right) \notag \\
 & -1.339( \pm 0.068) \log \left(M_{*} / 30\right) \notag \\
 & -1.601( \pm 0.055) \log \left(\frac{v_{\infty} / v_{\text {esc }}}{2.0}\right) \notag \\
 & +1.07( \pm 0.10) \log \left(T_{\text {eff }} / 20000\right) \notag \\
 & +0.85( \pm 0.10) \log \left(Z / Z_{\odot}\right),
\end{align}
where the ratio of the terminal flow velocity to the escape velocity ${v_{\infty} / v_{\text {esc}}}$ is determined from Galactic supergiants by \citet{1995ApJ...455..269L}, with ${v_{\infty} / v_{\text {esc}}} =2.6$ and ${v_{\infty} / v_{\text {esc}}} = 1.3$ in the Equations\ref{eq:Mdot_Vink_hot} and \ref{eq:Mdot_Vink_cool}, respectively \citep{2001A&A...369..574V, 2024arXiv240414488J}. The $L_{*}$ and $M_{*}$ are in solar units. For more details, please refer to the original literature.

When star evolves into the WR stage with $\rm{log}$ $(T_{\text {eff}}/K) > 4.0$ and $X_{s} \le 0.4$, the empirical prescription of mass loss provided by \citet{2000A&A...360..227N}. The original formulation shown as given in Equation \ref{eq:Mdot_WR}, the mass-loss rate depends strongly on both the luminosity ($L$) and the chemical composition.
 
\begin{align}
\label{eq:Mdot_WR}
\log \dot{M}= -11.00+1.29( \pm 0.14) \log L   \notag \\
 + 1.73( \pm 0.42) \log Y   \notag \\
 + 0.47( \pm 0.09) \log Z ,
\end{align}
where $Y$ and $Z$ represent the abundances of helium and heavier elements, respectively.

For all stars with lower effective temperatures where $\rm{log}$ ${(T_{\text{eff}}/K)} < 4.0$, we adopt the prescription proposed by \citet{1988A&AS...72..259D}, as given in Equation \ref{eq:Mdot_De}.

\begin{equation}
\label{eq:Mdot_De}
\log (-\dot{M})=1.769 \log L \\
-1.676 \log \left(T_{\text {eff }}\right)-8.158
\end{equation}

We find that increasing the stellar wind by the a factor of 1.5 allows most stars to evolve to the WNL stage with lower luminosities than by a factor of 1. This is more reasonable for explaining the low-luminosity Galactic samples. Thus, we adopt an overall scaling factor of $\eta_{\text {Dutch}} = 1.5$.

When very massive stars evolve to the red supergiant (RSG) phase, they may potentially approach the Eddington limit ($\Gamma_{\text {Edd}} \propto L/M$). According to \citet{2008A&A...482..945G}, \citet{2011A&A...531A.132V}, and \citet{2015MNRAS.452.1068C}, as recently suggested by \citet{2023A&A...678L...3V}, the mass-loss rates will increase when the stars approach the electron scattering Eddington limit $\Gamma_{\text {Edd}}$. The $\Gamma_{\text {Edd}}$ is defined as:

\begin{equation}
\Gamma_{\text {Edd}} = \frac{ L }{L_{\rm Edd}},
\label{eq:Edd}
\end{equation}
the ${L_{\rm Edd}}$ is the Eddington luminosity defined as:
\begin{equation}
L_{\text {Edd}} = \frac{4\pi cGM }{\kappa },
\label{eq:Ldd}
\end{equation}
where $\kappa$ represents the opacity, and $c$ is the speed of light. Thus when the stellar luminosity exceeds the $L_{\text {Edd}}$ by a factor of 5, mass loss is amplified by a factor of 3 artificially.

\subsection{Rotation} \label{subsec:rotation}
We initiate the initial surface rotation velocity in our rotating model at a value of ${v}/{v}_{\text {crit}}=0.4$ (here, $v$ is the initial surface rotation velocity at the equator), a ratio more suitable for describing stars of different initial masses than a fixed value.

In our study of rotation-induced instabilities, we account for the Spruit-Tayler dynamo (ST), dynamical shear instability (DSI), secular shear instability (SSI), Solberg-Høiland (SH) instability, Eddington-Sweet circulation (ES), and Goldreich-Schubert-Fricke (GSF) instability \citep{2000ApJ...528..368H, 2013ApJS..208....4P}. Ultimately, the contribution of rotation-induced instabilities to mixing can be calculated using the following formula:

\begin{equation}
D_{\text {mix}} =  D_{\text {mix(non-rot)}}+f_{\text c} *D_{\text {mix(rot)}},
\label{eq:Dmix}
\end{equation}
where the $D_{\text {mix(non-rot)}}$ represents the contribution of overshoot, convection, semiconvection, thermohaline, and other factors. $f_{\text c}$ is set to 1/30 according to \citet{2000ApJ...528..368H} and \citet{2016ApJ...823..102C}.

\begin{equation}
D_{\rm mix(rot)} =  D_{\rm DSI} + D_{\rm SH} + D_{\rm SSI} + D_{\rm ES} + D_{\rm GSF} + D_{\rm ST}.
\label{eq:Dnon-rot mix}
\end{equation}
The value of $D_{\rm mix(rot)}$ is determined by summing the products of all the instability parameters and factors described above.

The mass loss enhanced due to rotation is as follows:
\begin{equation}
\dot{M}(\Omega)=\dot{M}(0)\left(\frac{1}{1-\Omega / \Omega_{\text {\rm crit }}}\right)^{\xi}.
\end{equation}
The parameter $\dot{M}(0)$ is the mass-loss rate without rotation, $\xi$ is assumed to be 0.43 \citep{1998A&A...329..551L}. 
$\Omega_{\text {\rm crit}}$ is the critical angular velocity at surface:
\begin{equation}
\Omega_{\text {\rm crit}}^{2}=\left(1-\frac{L}{L_{\mathrm{Edd}}}\right) \frac{G M}{R^{3}}.
\label{eq:angular velocity}
\end{equation}

\subsection{Input} \label{subsec:input}
 
In this study, the composition fractions are adopted from the GS98 catalog \citep{1998SSRv...85..161G}, providing a comprehensive set of solar abundances for the stellar evolution calculations. The initial mass and metallicity for each model are specified according to the values listed in Table~\ref{tab:initial_parameter}.

The initial He mass fraction ($Y$) at varying metallicity is determined by the formula $Y=0.24+2Z$, where $Z$ is initial metallicity. For each given value of $Z$ corresponding a pair of $Y$ and H abundance ($X$). The terminal criterion of the program is established when the central He abundance ($Y_{\mathrm c}$) falls below $10^{-5}$. We divide the interior of each star into at least 10000 layers, and execute 66+66 stellar trajectory models for rotating and non-rotating stars calculated by the $k-\omega$ model, along with 66+66 tracks for the rotating and non-rotating stars calculated by the OV model, respectively.

\begin{deluxetable*}{cccccccccccc}
%\tablenum{1}
\tablecaption{Description of the model input.} \label{tab:initial_parameter}
\tablewidth{0pt}
\tablehead{
\colhead{Initial Parameters} & \multicolumn{8}{c}{Setting Values}}
%\decimalcolnumbers
\startdata
		$M_{\mathrm i}$/$\mathrm{M}_{\odot}$ & 50 & 60 & 70 & 80 & 90 & 100 & 110 & 120 & 130 & 140 & 150 \\
            $Z$ & 0.002 &  0.008  & 0.014 & 0.02 & 0.03  & 0.04  & \\ 
            $\eta_{\text {Dutch}}$  & 1.5\\
		${v}/{v}_{\text {crit}}$ &  0.4 &  0 &\\  
            $f_{\text {ov}}$ & 0.016 &  \\
\enddata
\end{deluxetable*}

\section{Results and Analysis} \label{sec:output}

\subsection{Hertzsprung–Russell Diagram} \label{sec:HR}
\begin{figure*}
\gridline{\fig{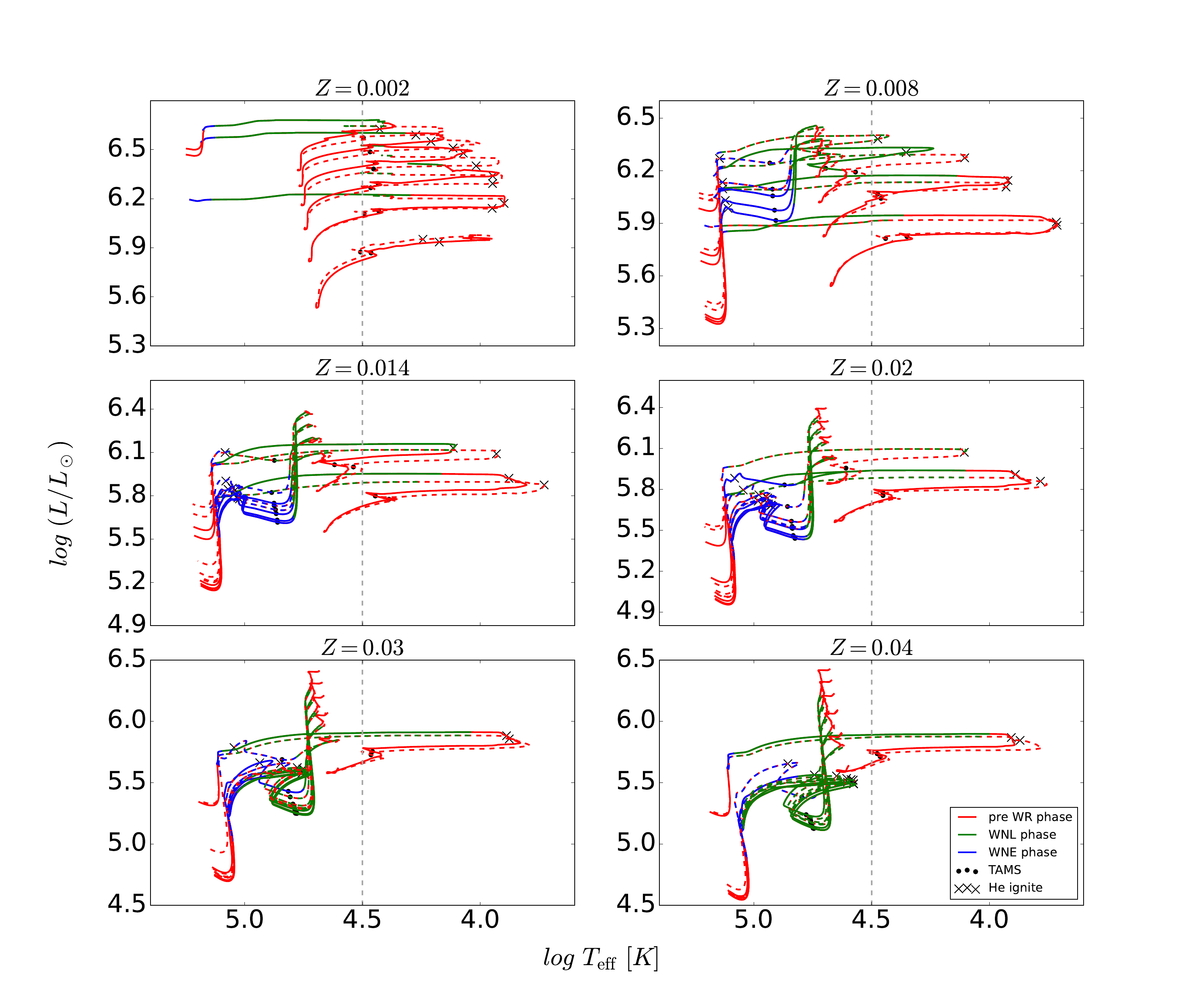}{0.49\textwidth}{(a)}
          \fig{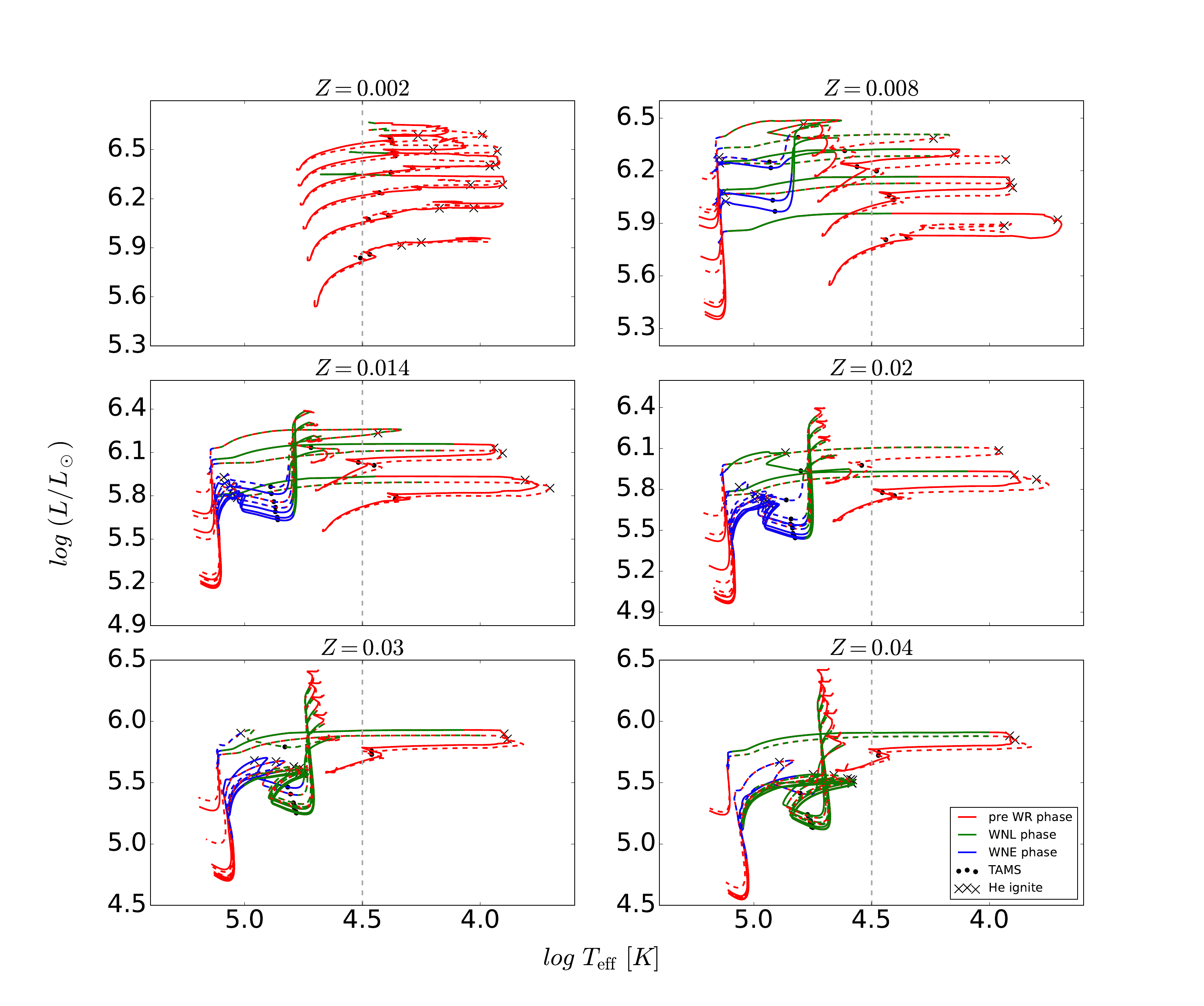}{0.49\textwidth}{(b)}    
          }
\caption{The HR diagram shows the evolutionary tracks of stars across various metallicities, encompassing initial masses from $50$ $\mathrm{M}_{\odot}$ to $150$ $ \mathrm{M}_{\odot}$ in increments of $\Delta\mathrm{M} = 20$ $ \mathrm{M}_{\odot}$. The panel (a) is rotating models with $v/v_{\rm crit} = 0.4$, the panel (b) is non-rotating models. The tracks are computed using both the KO model (solid lines) and the OV model (dotted lines). The vertical dashed lines are marked at the effective temperatures that satisfy $\log{(T_{\text{eff}}/K)}=4.5$, one of the criteria for classifying O-type stars taken from \citet{2012A&A...542A..29G}. The green sections of the trajectories signify the WNL phases. The blue portions indicate the WNE phases. Since other stages are not considered in this work, a consistent representation with red lines is used.
\label{fig:Figure(HR)}}
\end{figure*}

The evolutionary tracks of the rotating models computed using the KO and OV models, with initial masses ($M_{\mathrm i}$) of 50 to 150 $\mathrm{M}_{\odot}$ are presented in the pannel (a) of Figure~\ref{fig:Figure(HR)}.

In general, the Hertzsprung–Russell (HR) diagram reveals that massive stars with lower initial masses and metal-poor compositions tend to form the WNL stars later compared to those with higher initial masses and metal-rich compositions.
Conversely, these massive stars with higher metallicities exhibit a notable tendency to form the WNL earlier during the MS phase. 
For $Z \ge 0.014$ cases, most models calculated by the KO with higher initial masses ($M_{\mathrm i} \ge 70$) can form WNL stars before the terminal-age main sequence (TAMS) and tend to evolve into the lower luminosity region, even these higher-mass counterparts may form WNE stars before the ignition of He.
 
The effective temperature range of most WNL stars modeled by the OV is significantly smaller than that of those modeled by the KO at $Z \le 0.008$. Additionally, the luminosity of the metal-rich WNL stars ($Z \ge 0.02$) modeled by the KO undergoes a rapid decline within a narrow range of effective temperature, except for the $50$ $\mathrm{M}_{\odot}$ model, which shows a less pronounced effect. The luminosity shows significant variability, and the effective temperature shift blueward during the WNL phase. This is a consequence of the increasing display of H-burning products on the surface near the end of the MS stage, caused by the mixing induced by convection and rotation, which promotes the strength of mass loss.
  
Similarly, non-rotating models are employed for comparative analysis, as depicted in the pannel (b) of Figure~\ref{fig:Figure(HR)}. The WNL stars formed by the KO model show a higher likelihood of occurring under the same initial conditions compared to those formed by the OV model, especially noticeable under lower metallicity ($Z = 0.002$). Compare the two cases based on the Figure~\ref{fig:Figure(HR)}, rotation induces more pronounced evolutionary discrepancies in the formation of the metal-poor WNL stars, especially in the OV models. This behavior aligns with the finding of \citet{2020MNRAS.494.3861R}, where rotation at [Fe/H] = -1 and -2, with $v/v_{\text {crit}} = 0.4$ and $0.3$, respectively, drives the evolution of O-type star towards the WNL phase. From the point of the mixing, a plausible explanation is that the mixing region of nucleosynthesis products for rotating star is more extensive compared to non-rotating star, thereby reducing the thickness of the outer H-rich layers and extending the mixing region. This enables the stellar surface to achieve WNL criteria at an accelerated rate.

\subsection{Mass and lifetime} \label{sec:Mass and lifetime}
\subsubsection{Mass distribution} \label{sec:mass} 
\begin{figure*}
\gridline{\fig{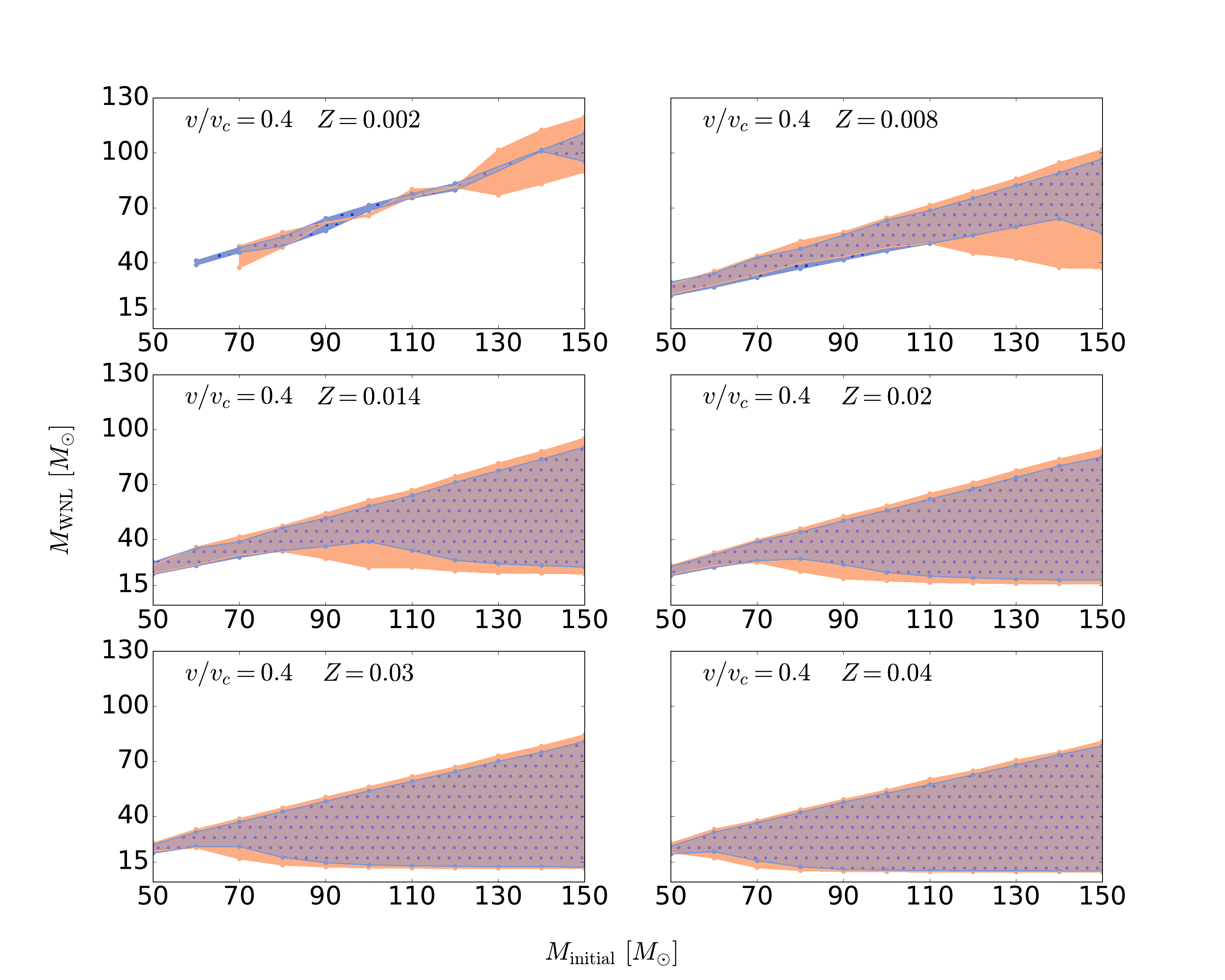}{0.49\textwidth}{(a)}
          \fig{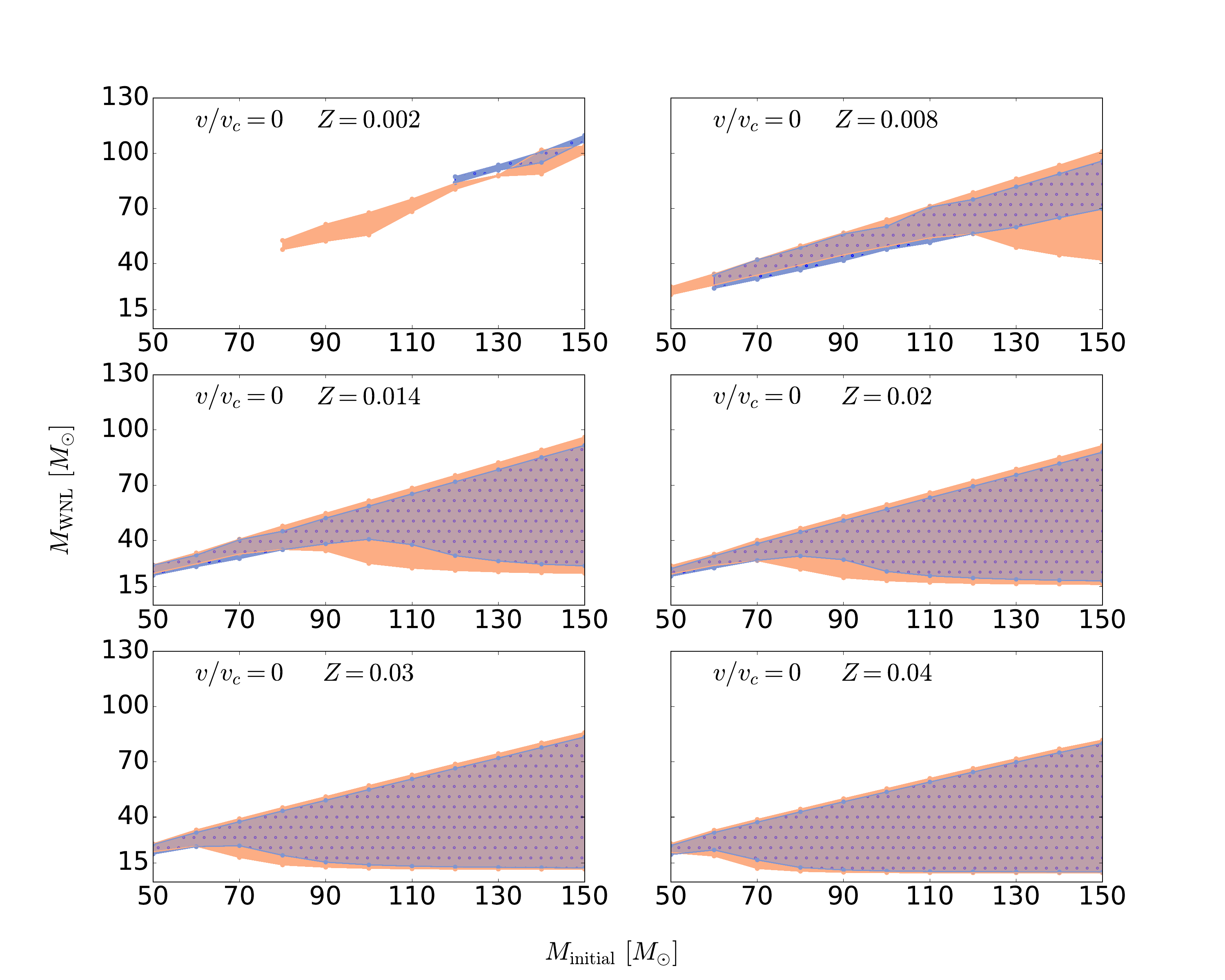}{0.49\textwidth}{(b)}      
          }
\caption{Statistics of mass ranges for stars during the WNL stage with various metallicities and initial masses. The orange part represents the results of KO, and the blue part represents the results of OV. The panel (a) corresponds to a surface rotational velocity of $v/v_{\text {crit}} = 0.4$, while the panel (b) provides results for a rotational velocity satisfying $v/v_{\text {crit}} = 0$.
\label{fig:Figure(KO & OV im & WNL_mass)}} 
\end{figure*}

Figure~\ref{fig:Figure(KO & OV im & WNL_mass)} presents the mass range of the WNL stars obtained using the two overshooting schemes under rotating and non-rotating conditions, as illustrated in panels (a) and (b), respectively.
 
In general, all models exhibit a common trend: The lower initial masses and lower terminal masses of the WNL stars correspond to higher-metallicity models. This trend becomes more pronounced with the increasing Z due to the exponential relationship between the mass loss rate and the metallicity. Generally, the models with lower initial masses and lower metallicities exhibit a narrower range of masses during the WNL phase. In contrast, the metal-rich WNL stars ($Z \ge 0.02$) with higher initial model masses handled by the two overshooting models exhibit similar terminal masses. As detailed in the Appendix (Table~\ref{KO} and Table~\ref{OV}).

In detail, the OV model generally predicts lower initial masses and slightly higher terminal masses for the WNL stars compared to those predicted by the KO model, under the same initial conditions. 
In other words, the results from the OV model indicate that the evolutionary mass range of WNL stars is generally smaller than that of KO in most cases, especially for metal-poor massive stars. The mass ranges of rotating WNL stars treated by the two overshooting schemes in all models are as follows: for the KO model, 21-95 $\mathrm{M}{\odot}$ at Z=0.014 and 16-89 $\mathrm{M}{\odot}$ at $Z=0.02$; for the OV model, 21-91 $\mathrm{M}{\odot}$ at Z=0.014 and 18-85 $\mathrm{M}{\odot}$ at $Z=0.02$. The reasons for this distribution will be explained in the next parts (Section~\ref{sec:convection and convective overshooting} and Section~\ref{sec:Overshooting}).

Furthermore, it can be clearly seen that the lowest model limit for WNL formation obtained from the two overshooting schemes is similar when considering rotation. Both the KO and the OV models can evolve into WNL phases at $Z = 0.002$, $M_{\mathrm i} = 70$ $\mathrm{M}_{\odot}$ and $M_{\mathrm i} = 60$ $\mathrm{M}_{\odot}$, respectively. While in the non-rotating case, a higher formation criterion is observed at $Z=0.002$, an initial mass of $M_{\mathrm i} = 80$ $\mathrm{M}_{\odot}$ for the KO model, and $M_{\mathrm i} = 120$ $\mathrm{M}_{\odot}$ for the OV model, respectively. This implies that rotation has a certain influence on the OV treatment of convective overshooting in metal-poor massive stars. 
 
In essence, in addition to the treatment of rotation and mass loss relay on the metallicity, the method to handle the overshoot of stellar interior plays a crucial role in shaping the evolutionary path and pattern of massive stars. Effective convective overshooting promotes the transport of nucleosynthesis products and the expansion of the mixing region, thereby enhancing the likelihood of WNL occurrence during the stellar evolution.

\subsubsection{Mass of convective core} \label{sec:convection and convective overshooting}
 
\begin{figure*} 
\plotone{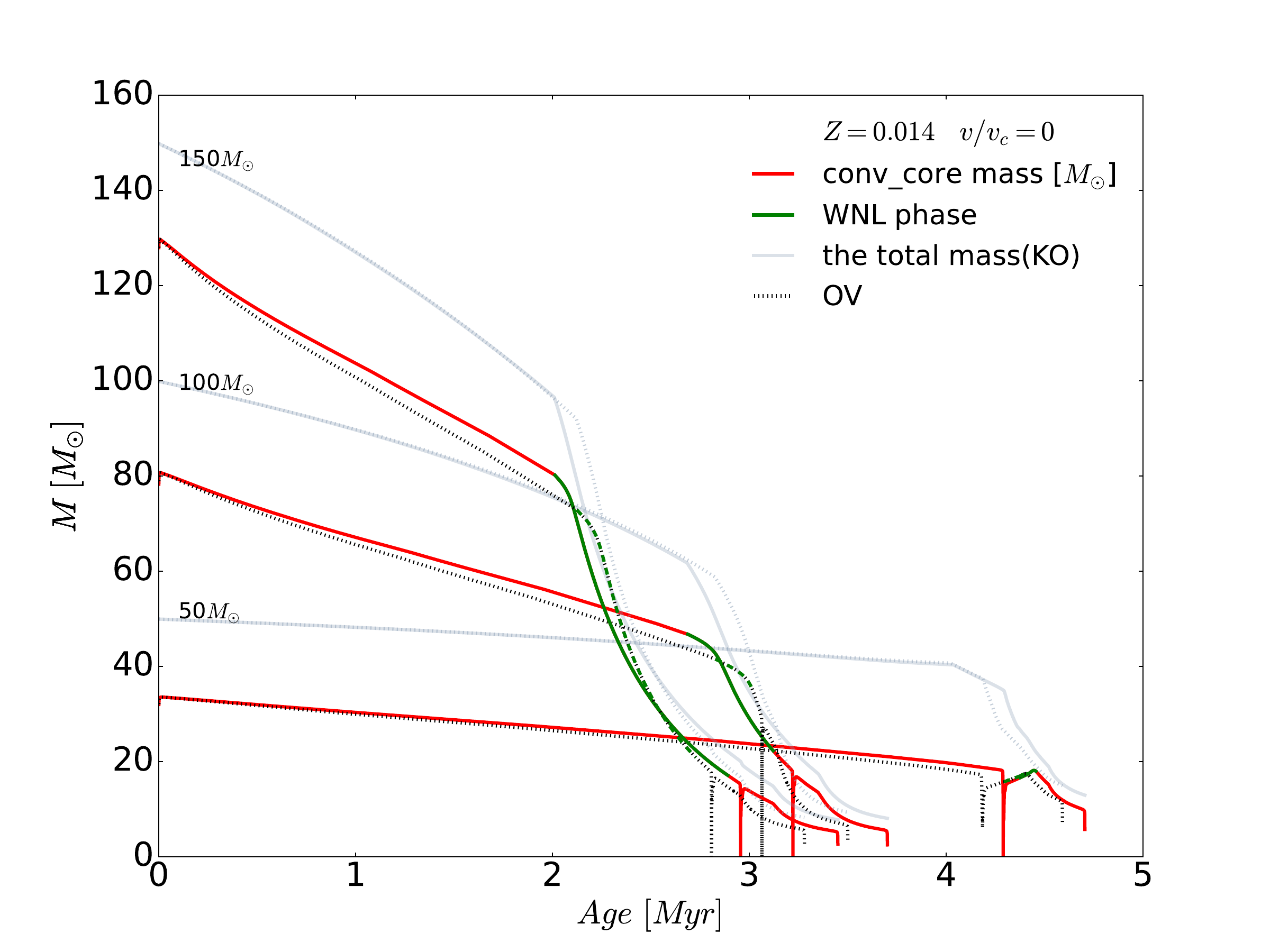}
\caption{The mass of the convective core of the non-rotating star is shown as a function of lifetime for different initial masses using both the KO and the OV models. The colored lines, arranged from bottom to top, illustrate the convective core masses corresponding to initial stellar masses of $50$ $\mathrm{M}_{\odot}$, $100$ $ \mathrm{M}_{\odot}$, and $150$ $\mathrm{M}_{\odot}$ at $Z=0.014$, respectively. The solid line represents the KO model, while the dotted line represents the OV model. The total mass is depicted with lower opacity. The green section on the evolutionary line of convective core mass represents the WNL stage.
\label{fig:Figure(conv core mass(0))}}
\end{figure*}

In order to assess the impact of various overshooting schemes treated by the KO and the OV models on the internal structure of WNL stars, Figure~\ref{fig:Figure(conv core mass(0))} shows the evolution of the convective core over the timescale calculated by the non-rotating models at $Z = 0.014$. A noticeable correlation is observed that the mass of the convective core increases as the initial mass increases. The results from the OV model consistently exhibit a slightly smaller size and shorter timescale of the convective core compared to the KO model throughout their evolutionary stages.  
 
1. In general, the emergence of WNL stars predicted by the KO model (depicted by the green solid line segment) occurs earlier than that predicted by the OV model during the H-burning phase (for $M_{\mathrm i} = 100$ $\mathrm{M}_{\odot}$ and $M_{\mathrm i} = 150$ $\mathrm{M}_{\odot}$ models. However, it occurs later than the OV model during the He-burning phase (for $M_{\mathrm i} = 50$ $\mathrm{M}_{\odot}$ model).  
 
2. Before evolving to the WNL stage, the convective core calculated by the KO model is larger than that calculated by the OV model.  
Thus, upon transitioning to the WR stage, the WNL stars modeled by the KO model form earlier and experience less loss of the H-rich envelope compared to those modeled by the OV model. This leads to a slightly higher total mass of the WNL star than the OV counterpart during the early WNL phase (for models with $M_{\mathrm i} = 100$ $\mathrm{M}_{\odot}$ and $M_{\mathrm i} = 150$ $\mathrm{M}_{\odot}$). 
This discrepancy explains why Figure~\ref{fig:Figure(KO & OV im & WNL_mass)} displays higher initial masses of WNL phases in most cases computed by the KO. 
Additionally, the size of the convective core contracts earlier in the KO model. This contraction is attributed to the enhanced mass loss as He and heavier elements increase on the surface during the late stage of the MS. This is also related to the transformation of the stellar wind formula into \citet{2000A&A...360..227N}, as established in the study by \citet{sander_vink_higgins_shenar_hamann_todt_2020}, which indicates that the \citet{2000A&A...360..227N} model exhibits higher levels of mass loss.

3. It is observed that the TAMS of the OV model occurs earlier than that of the KO model, consequently initiating He-ignition earlier. This results in a slightly larger convective core mass developing for the WNL stage during the core He-burning phase in the OV model ($M_{\mathrm i} = 50$ $\mathrm{M}_{\odot}$). Therefore, for lower-mass models evolving into WNL during the early He-burning phase, WNL stars modeled by the OV model tend to show higher initial masses and younger ages compared to those caculated by the KO model, particularly in cases of lower initial masses and/or low metallicities. This difference explains why the initial masses of some WNL stars in the OV model (most of which occur at $Z = 0.002$) in Figure~\ref{fig:Figure(KO & OV im & WNL_mass)} may be higher than that in the KO model, suggesting that these stars may be formed during the core He-burning phase.

\subsubsection{Impact of Overshooting on evolution} \label{sec:Overshooting} 
 
\begin{figure*}       
\gridline{\fig{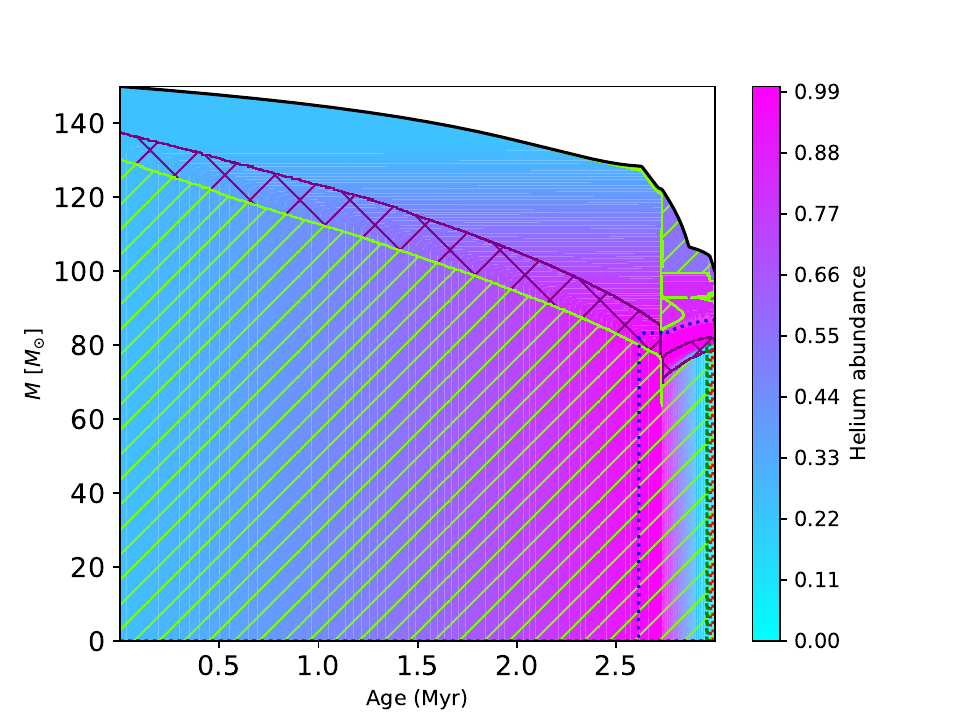}{0.49\textwidth}{(a)}
          \fig{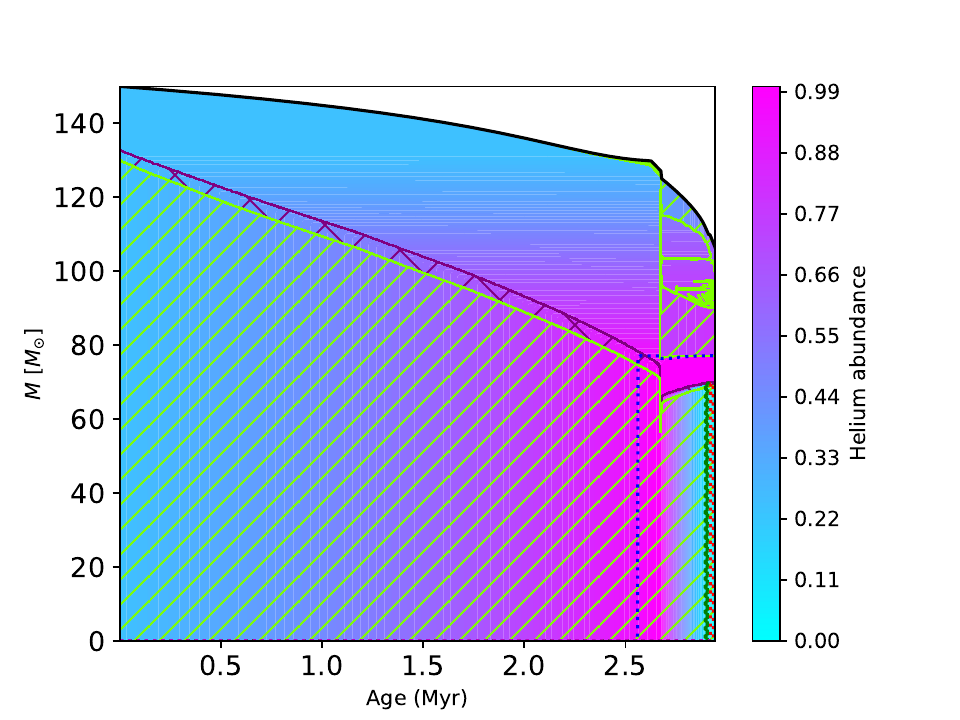}{0.49\textwidth}{(b)}      
          }
\caption{Kippenhahn diagram for the non-rotating model with $M_{\mathrm i} = 150$ $\mathrm{M}_{\odot}$ and $Z = 0.002$. Panel (a) displays the result computed using the KO model, while panel (b) shows the result from the OV model. The green hatched lines indicate the convective region, the purple hatched lines represent the overshoot region, and the blue, red, and green dotted lines correspond to the masses of the He core, the C core, and the O core, respectively.
\label{fig:kipp}}
\end{figure*}

In addition to the fact that convection correlates with the mass of the WNL star, overshooting is also significantly associated with the evolution of the WNL star, as core overshooting not only affects the size of the convective core but also influences the width of the MS band \citep{2023ApJS..268...51L}. The overshoot mixing and mass loss are the main factors leading to the anomalous surface enrichment of the WNL stars. Consequently, the masses and lifetimes of the WNL stars are determined by the extension of the mixing regions. We find that the previous results of the KO and OV models show significant differences for low-metallicity ($Z=0.002$) when utilizing the two overshooting schemes. The Kippenhahn diagrams of $150$ $\mathrm{M}_{\odot}$ and $Z = 0.002$ non-rotating models with the overshoot mixing schemes of the KO model and the OV are shown in Figure~\ref{fig:kipp}.

The Kippenhahn diagrams show an obvious transition in the width of the overshooting region between the MS and post-MS stages. Additionally, it can be observed that the overshooting region in the KO model is notably wider than that in the OV model, with this difference being more pronounced during the MS stage. This facilitates the outward expansion of the H-burning zone, thereby extending the lifetime of the MS and increasing the mass of the He-core at the TAMS. However, the discrepancy in the overshoot mixing region between the two models gradually insignificant during the post-MS stage. This reduction is a result of the combined effects of mass loss and overshoot mixing during the late stage of the MS.

The Figure~\ref{fig:kipp} explains why, in the HR diagram, the evolution of most stars with higher initial masses and metallicities ($Z \ge 0.014$) in the KO model shifts towards higher temperatures soon after the onset of the MS. This shift is due to efficient mixing through convection and overshooting, which exposes more H-burning products on the surface, reduces opacity, and enhances the strength of mass loss. Consequently, this leads to a larger chemically homogeneous zone and rapid stripping of the H-rich envelope.

The discrepancy between the two overshooting schemes provides a foundational rationale for tending to use the $k-\omega$ model. However, it's crucial to note that due to observational constraints, the KO model may not necessarily offer the optimal explanation for observed phenomena.

\subsubsection{lifetime and ratio} \label{sec:Age and duration}

\begin{figure*}
%\vspace{-10pt} 
\plotone{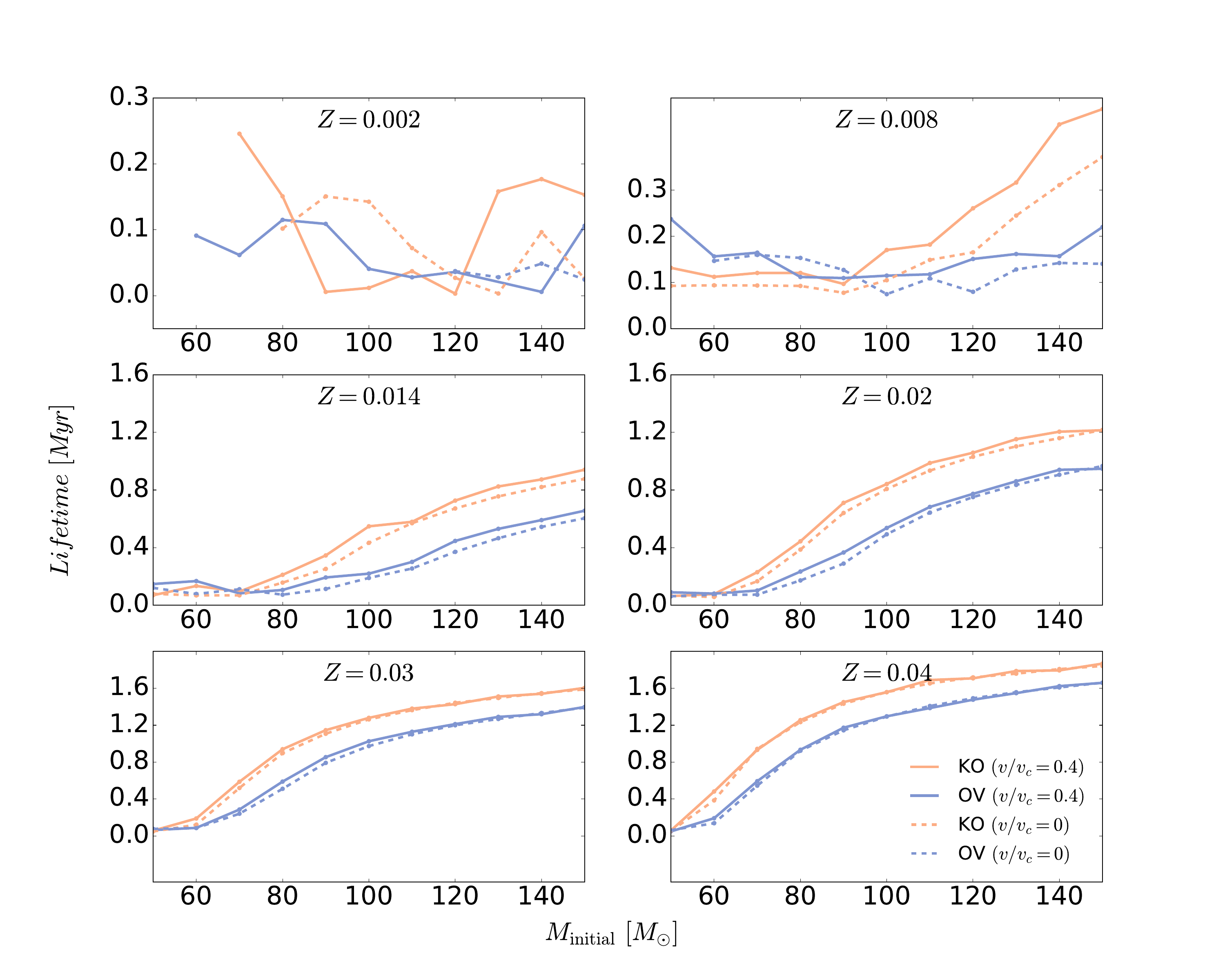} 
\caption{Lifetime distribution of the WNL stars for different initial mass models. The orange and blue lines represent the KO and the OV models, respectively, while the solid and dashed lines indicate the presence or absence of initial surface rotational velocities.
\label{fig:Figure(KO & OV im & WNL_lifetime 0.4 & 0)}}
\end{figure*}

\begin{figure*} 
\centering
\plotone{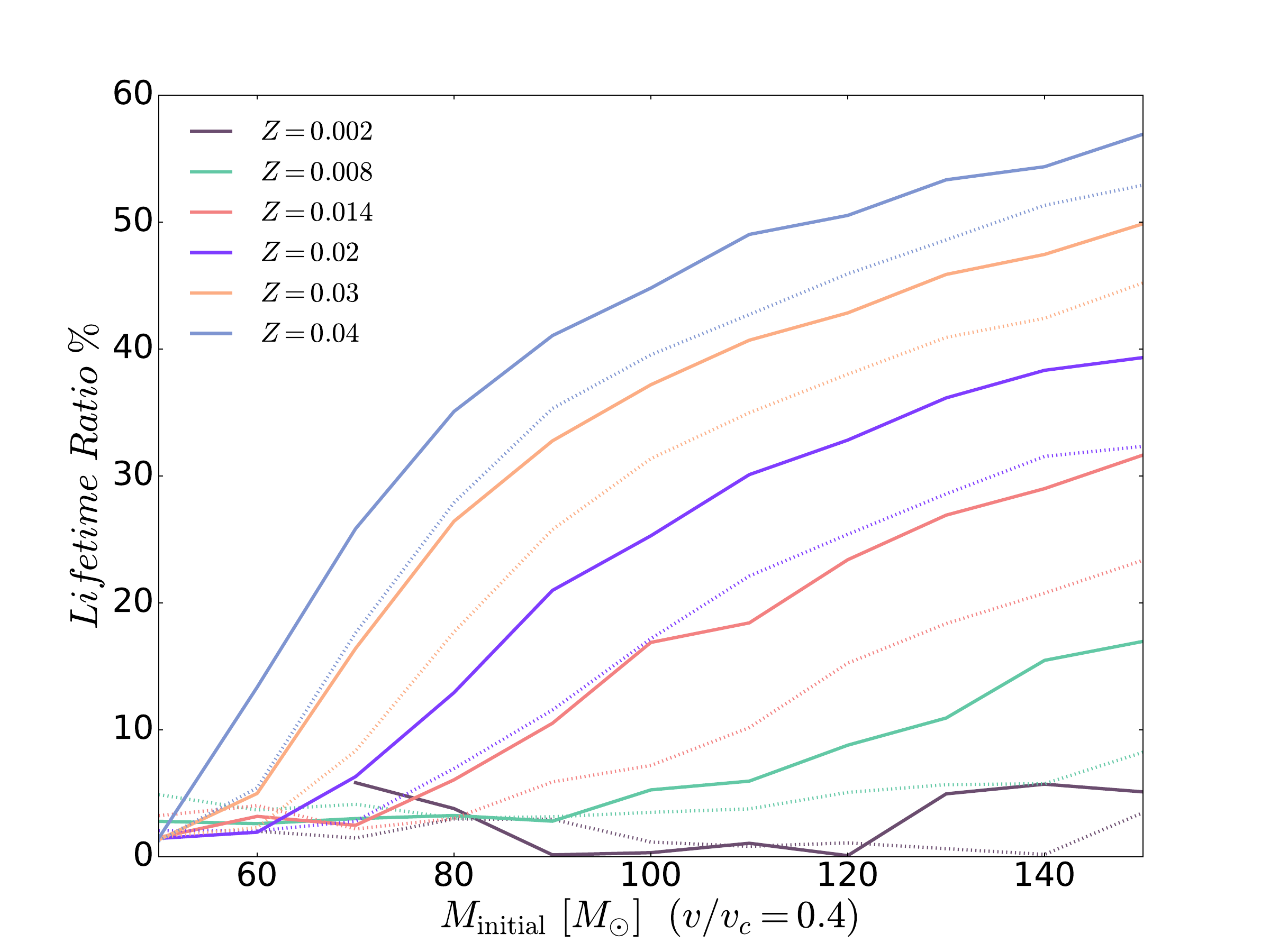} 
\caption{Distribution of the proportional lifetimes of WNL stars relative to their total lifetimes at the moment of WNL termination, as obtained from the rotating KO and OV models. The solid lines represent the KO model, and the dashed lines represent the OV model.
\label{fig:Figure(lifetime percent)}}
\end{figure*}

The lifetimes of the WNL stage and the proportions of this stage in relation to the total lifetime are presented in Figure~\ref{fig:Figure(KO & OV im & WNL_lifetime 0.4 & 0)} and Figure~\ref{fig:Figure(lifetime percent)}, respectively. The moment of WNL termination is recorded as the total lifetime.
Based on the Figure~\ref{fig:Figure(KO & OV im & WNL_lifetime 0.4 & 0)}, the following features are observed:

1. The two various overshoot mixing schemes significantly influence the lifetime of the WNL phase. For most cases of $Z \ge 0.008$ models, the WNL star handled by the KO model possess a longer lifetime than those handled by the OV model. This difference is due to the KO model leading to a broader mixing region compared to the OV treatment.
At $Z \ge 0.014$, as metallicity increases, the uncertainties in lifetimes due to metallicity and rotation become less significant, being instead dominated by the uncertainty caused by the overshoot mixing.

2. Except for the metal-poor cases, the lifetimes of WNL stars increase with the increasing initial model masses at the same metallicity.

3. At $Z = 0.002$, there is no significant correlation observed between the lifetimes of the metal-poor WNL stars and the initial model masses across various overshooting schemes and rotation conditions.
 
4. At $Z \ge 0.014$, rotating WNL stars generally have a slightly longer lifetime compared to non-rotating counterparts in both the KO and the OV models, though the impact of rotation becomes less pronounced compared to the influence of the different overshooting schemes. This is because metal-rich and massive counterparts show a more pronounced impact on the transfer of angular momentum. In higher metallicity scenarios, mass loss is more pronounced than in lower metallicity scenarios, leading to stellar contraction to maintain stability. More angular momentum is transferred outward from the contracting core and carried away by the lost material, resulting in a faster decline in surface equatorial velocities. As a result, the impact of surface velocities in metal-rich stars with higher initial masses becomes comparable to that in non-rotating cases during the late stage of the MS.

In the Figure~\ref{fig:Figure(lifetime percent)}: For metal-rich stars ($Z = 0.03$) with high initial masses, a significant portion of their evolutionary timescales may be spent in the WNL phase, constituting nearly $50\%$ of the total duration (for the KO model with $M_{\mathrm i} = 150$ $\mathrm{M}{\odot}$, refer to the Appendix for more details). Thus, during the short lifetimes of the metal-rich massive stars, a significant proportion will be in the WNL stage. At $Z \le 0.014$, the WNL stars modeled by the KO occupy less than $32\%$ of their total lifetimes, even with a sufficiently large initial mass ($150$ $\mathrm{M}{\odot}$ model). The WNL samples account for about 30\% of the total WR stars according to the statistics of \citet{2023ApJS..268...51L}. This suggests that when using the KO model with an initial mass of $150$ $\mathrm{M}{\odot}$ to simulate the evolution of the galactic WR population, perhaps the metallicity of $Z = 0.014$ may be more accurate. Similarly, for the OV model, the metallicity of $Z = 0.02$ is likely closer to the actual value. In fact, only a few WNL stars evolve from massive stars with an initial mass of $150$ $\mathrm{M}{\odot}$. Considering that the main sequence accounts for more than $90\%$ of stellar evolution, it can be inferred that a significant proportion of high-mass stars is likely to be observed as WNL stars, particularly in metal-rich environments.

\subsection{Distribution of Surface Element Abundance} \label{sec:Elemental abundance}

\begin{figure*}
\centering
\includegraphics[width=1\linewidth]{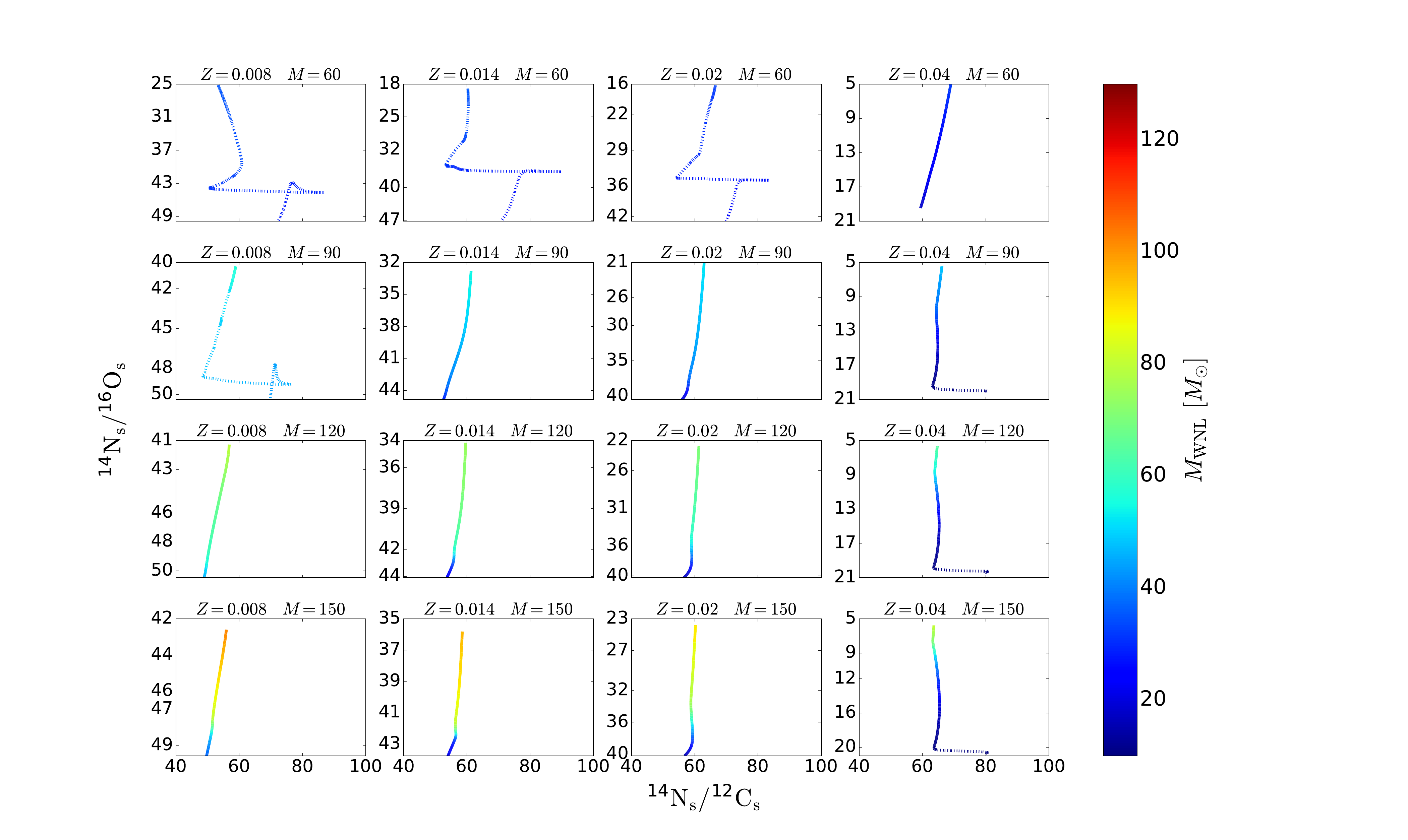}  
\caption{The evolution of relative surface mass abundance ratios during the WNL stage simulated by the KO model with rotation is depicted. The solid line represents the core H-burning phase, while the dashed line represents the core He-burning phase. The colored lines indicate the masses of the WNL stars.
\label{fig:Figure(CNO ratio)}}
\end{figure*}

It is widely accepted that massive stars display surface enrichment in elements due to mass loss and mixing. According to \citet{2023ApJS..268...51L}, the extended overshooting region can work with the rotational mixing to explain the significant fraction of the nitrogen enrichment. Previous sections have demonstrated that most metal-rich stars with higher initial masses are more likely to form WNL stars before core H exhaustion. In contrast, WNL stars with lower metallicity and/or lower initial mass are more likely to form during the core He-burning phase. We categorize these WNL stars into two categories and analyze the distribution of surface abundance ratios for the rotating models treated with both the KO and the OV schemes, respectively. One is those that evolve into WNL during the H-burning phase (abbreviated as msWNL), while the other is those that evolve into WNL during the He-burning phase (abbreviated as heWNL). It is important to note that these designations for WNL stars are distinct from the previously mentioned WNh and classic WNL stars. They serve solely as a convenient way to distinguish whether the WNL star forms during the core H- or He-burning phases and are not related to the previous spectroscopic classifications of subclasses. The surface mass abundance ratios of N/C and N/O during the WNL stage (note: this study utilizes the isotopes $^{12}$C, $^{14}$N, and $^{16}$O for the ratio calculations) are shown in Figure~\ref{fig:Figure(CNO ratio)}. 

1. The metal-rich WNL stars with higher initial masses primarily undergo H-burning, whereas the metal-poor WNL stars with lower initial masses predominantly undergo He-burning.

2. The value of N/O increases monotonically with mass loss during the core H-burning stage, and the evolutionary range of metal-rich WNL stars occupies a slightly wider range than their metal-poor counterparts due to mass loss, but the value of N/C does not change significantly during this stage.

3. The value of N/C increases and then decreases during the core He-burning stage.

The reasons for this trend are explained as follows:

As we know that the CNO cycle is the primary mechanism for energy production in massive stars during the core H-burning phase, which can be further divided into the CN-cycle and NO-cycle. Due to the considerably higher reaction rate of the CN-cycle compared to the NO-cycle, the CN-cycle reaches equilibrium swiftly within the order of ten thousand years, facilitating the conversion of $^{12}$C into $^{14}$N. Furthermore, the $^{14}$N generated in the NO-cycle can act as reactants for the CN-cycle, further enhancing its efficiency. Consequently, during the MS phase, the $^{12}$C abundance is already within a stable range. Therefore, the value of N/C should be relatively constant, and the amount of N/O will change significantly during the msWNL phase (solid line).

Additionally, due to the rapid equilibrium reached by the CN-cycle in low-metallicity massive stars compared to those with higher metallicity, take $Z=0.03$ models for example, the CN-cycle reaches equilibrium in less than 10,000 years, indicating an even earlier equilibrium at $Z=0.008$. Similarly, substantial differences exist in the timescales of $^{16}$O consumption at different chemical compositions. This is due to the prolonged duration of more $^{16}$O depletion in high-metallicity stars compared to low-metallicity stars. And the inefficient NO-cycle may also not reach equilibrium by the core H exhaustion for those metal-rich massive stars. Generally, as a result, for the msWNL stars, metal-rich counterparts enter the msWNL stage before $^{16}$O is completely exhausted due to their earlier onset, while metal-poor counterparts may approach near-complete exhaustion of $^{16}$O due to the later onset and shorter duration of the msWNL phase. Consequently, we can infer that high-metallicity stars entering the msWNL phase, the $^{14}$N on the surface primarily originates from the CN-cycle and partially from the non-equilibrium NO-cycle. In contrast, for low-metallicity msWNL stars, surface $^{14}$N is entirely derived from the CN- and NO-cycles. This allows for speculation on the specific stage of stellar evolution from which WNL stars originate. 

Furthermore, the evolution of heWNL stars is more complex. Metal-poor and/or lower-mass massive stars tend to form the WNL stars during the core He-burning phase, where N is primarily produced in the He-burning phase by the diffusion of $^{12}$C into the H-burning shell through rotation \citep{2002A&A...390..561M} and overshoot. However, as more $^{12}$C gradually increases and is transported to the surface due to mass loss and mixing, the trend of the N/C ratio shows an initial increase followed by a decrease. As a result, the N/C ratio exhibits a more intricate range of variations. 

\begin{figure*} 
\centering 
\includegraphics[width=1\linewidth]{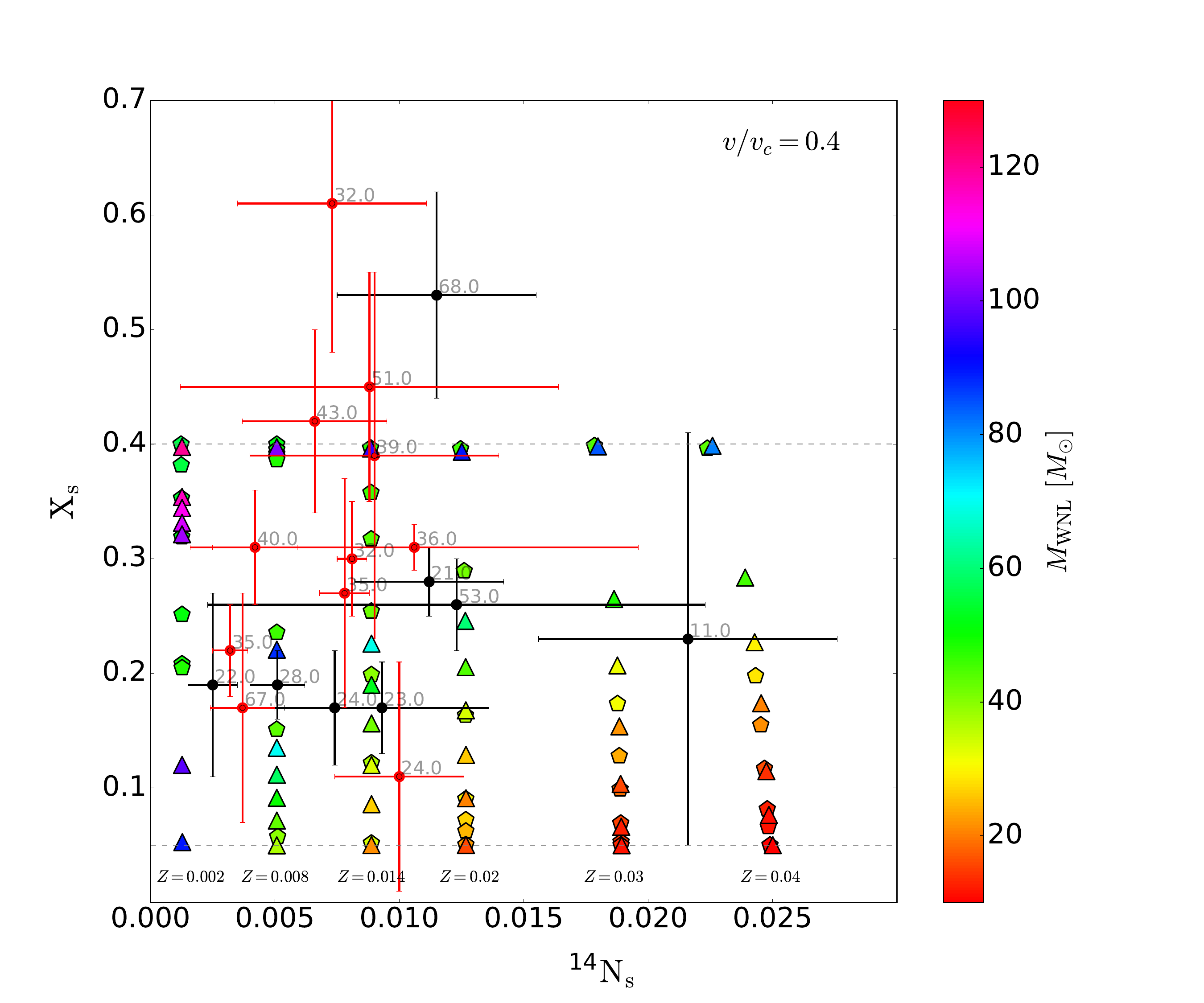} 
\caption{Surface H mass fraction as a function of surface N mass fraction for the rotating WNL stars. The black and red filled dots represent the WNL stars in the MW and the LMC, respectively, based on the observed surface abundance data from \citet{2023A&A...680A..22M}. The numbers indicate the masses of the WNL stars, as referenced in \citet{201962557Hamann} and \citet{201456527Hainich}, while the colored triangles depict the distribution of surface elements during the WNL phase from the KO model. These distributions are obtained through simulations with different initial metallicities. The columns, from left to right, represent $Z$ of 0.002, 0.008, 0.01, 0.014, and 0.02, with an initial mass of $150$ $\mathrm{M}{\odot}$, respectively. Similar, the pentagons indicate the $80$ $\mathrm{M}_{\odot}$ models. The filled colors reflect the masses of the modeled WNL stars.
\label{fig:Figure(N-H model & obs)}}
\end{figure*}

For comparison, we study the surface mass abundance and error ranges of the H, He, C, and N for eight samples from the Milky Way (referred to as MW-WNL) and eleven samples from the Large Magellanic Cloud (referred to as LMC-WNL), as listed in the latest literature by \citet{2023A&A...680A..22M}. In Figure~\ref{fig:Figure(N-H model & obs)}, we mark the range of surface H abundance between 0.05 and 0.4 with dashed lines, indicating four samples with surface H levels significantly exceeding the threshold of 0.4. We focus solely on the distribution of surface elements for WNL stars with different initial masses, without considering the consistency of metallicities for samples with the models. Additionally, it appears from the figure that the WNL samples from the MW and the LMC roughly correspond to the models with metallicities ranging from $Z = 0.002$ to $Z = 0.02$ that we have adopted. Overall, when using the model with an initial mass of $150$ $\mathrm{M}_{\odot}$, it seems to predict a higher mass than the actual observed values, while the $80$ $\mathrm{M}_{\odot}$ model appears to perform better.  

\begin{figure*} 
\centering    
\includegraphics[width=1\linewidth]{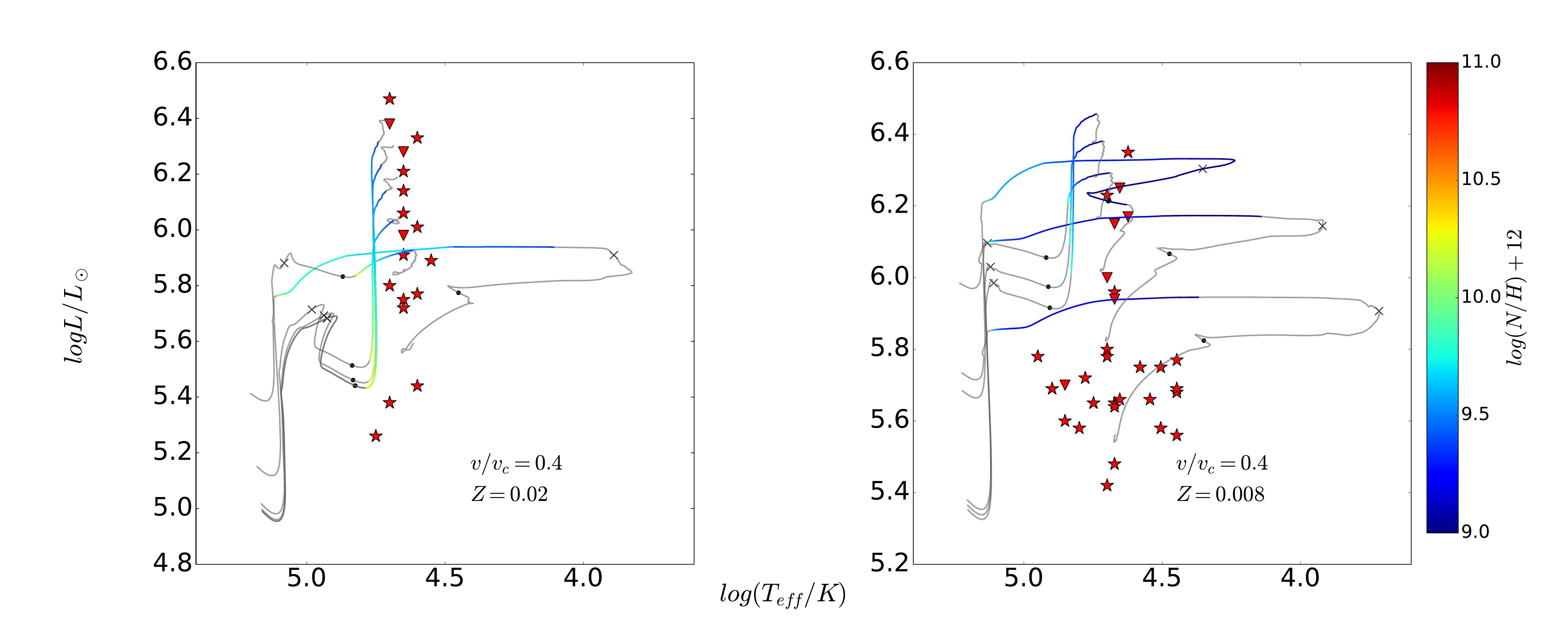}  
\caption{Evolutionary tracks in the HR diagram for the rotating models handled by the KO model. Each track is caculated with its initial mass from $50$ $ \mathrm{M}_{\odot}$ to $150$ $ \mathrm{M}_{\odot}$ in increments of $20$ $ \mathrm{M}_{\odot}$. The observed MW and LMC samples from \citet{201962557Hamann} and \cite{201456527Hainich}, repectively. The colored lines correspond to the surface number abundance of N on a logarithmic scale where the abundance of H is 12. Asterisks represent single WNL stars, while triangle symbols indicate WNL stars that are classified as binaries or are considered binary suspects. 
\label{fig:Figure(Comparison)}}
\end{figure*}

\begin{deluxetable*}{ccccccccc}
%\tablenum{}
\tablecaption{Description of the MW-WNL stars.\label{tab:MW-WNL}}
\tablewidth{0pt}
\tablehead{
\colhead{Name} &    \colhead{Spectral subtype} &    \colhead{WNL} &   \colhead{$\rm{log}$ $L$} &   \colhead{$\rm{log}$ $T_*$} &   \colhead{$X_{\rm H}$} &   \colhead{$M$} &    \colhead{$\rm{log}$ $\dot{M}$}&   \colhead{$M_{\text {WR }}$}   \\
 \colhead{WR} &    \colhead{ } &  \colhead{S/B} &   \colhead{ $(L_{\odot})$}  &    \colhead{(K)} &    \colhead{${[\%]}$} &    \colhead{$(M_{\odot}$)} &    \colhead{($M_{\odot}*yr^{-1}$)} &    \colhead{$( M_{\odot}$)} 
}
\decimalcolnumbers
\startdata
12  & WN8h + OB        & B   & 5.98 & 4.65  & 27   & 31/30  &  -4.3    & 31    \\
16  & WN8h             & S   & 5.72 & 4.65  & 25   & 21/19  &   -4.6    & 21    \\
22  & WN7h + O9III-V   & B   & 6.28 & 4.65  & 44   & 49/75  &   -4.4    & 49    \\
24  & WN6ha-w (WNL)    & S   & 6.47 & 4.70  & 44   & 68/114  &   -4.3    & 68    \\
25  & WN6h-w+O (WNL)   & B   & 6.38 & 4.70  & 53   & 58/98   &   -4.6    & 58    \\
40  & WN8h             & S   & 5.91 & 4.65  & 23   & 28/26  &   -4.2    & 28    \\
78  & WN7h             & S   & 5.80 & 4.70  & 11   & 24/22  &   -4.5    & 24    \\
82  & WN7(h)           & S   & 5.26 & 4.75  & 20   & 11      &   -4.8    & 11    \\
85  & WN6h-w (WNL)     & S   & 5.38 & 4.70  & 40   & 13      &   -5.0     & 13    \\
87  & WN7h             & S   & 6.21 & 4.65  & 40   & 44/59  &   -4.5    & 44    \\
89  & WN8h             & S   & 6.33 & 4.60  & 20   & 53/87  &   -4.4    & 53    \\
105 & WN9h             & S   & 5.89 & 4.55  & 17   & 27/25  &   -4.4    & 27    \\
108 & WN9h             & S   & 5.77 & 4.60  & 27   & 23/21   &   -4.9    & 23    \\
116 & WN8h             & S   & 5.44 & 4.60  & 10   & 14      &   -4.4    & 14    \\
124 & WN8h             & S   & 5.75 & 4.65  & 13   & 22/20  &   -4.3    & 22    \\
131 & WN7h             & S   & 6.14 & 4.65  & 20   & 39 /44 &   -4.5    & 39    \\
156 & WN8h             & S   & 6.01 & 4.60  & 27   & 32 /32  &   -4.6    & 32    \\
158 & WN7h + Be?       & S   & 6.06 & 4.65  & 30   & 35/35   &   -4.7    & 35     \\
\enddata
\tablecomments{``M'' indicates the current stellar mass derived from the M-L relation for homogeneous He stars. If a second value is given, the latter is derived from an M-L relation for the WNL stars based on evolutionary tracks. We use the latter ($M_{\text {WR}}$) for comparison with the models. ``$X_{\rm H}$'' represents the surface H mass fraction, expressed in percentage terms. ``S'' represents the WNL star as a single star, and ``B'' represents the WNL star as a member of a binary system. Controversial stars are still represented by a single star. Consistent with the representation in Table~\ref{tab:LMC-WNL} below. The catalog of HMW18 is extracted from Table 1 and Figure 2 in the \cite{201962557Hamann}.}
\end{deluxetable*}
 
\begin{deluxetable*}{cccccccc}
\tablecaption{Description of the LMC-WNL stars.\label{tab:LMC-WNL}}
\tablewidth{0pt}
\tablehead{
\colhead{Name} &    \colhead{Spectral subtype} & \colhead{WNL} &   \colhead{$\rm{log}$ $L$} &   \colhead{$\rm{log}$ $T_*$} &   \colhead{$X_{\mathrm H}$} &   \colhead{$M$} &    \colhead{$\rm{log}$ $\dot{M}$}  \\ 
\colhead{BAT99} &    \colhead{ } &    \colhead{S/B} &   \colhead{ $L$$_{\odot}$}  &    \colhead{(K)} &    \colhead{ } &    \colhead{$M$$_{\odot}$} &     \colhead{[$M$$_{\odot}*yr^{-1}$]} 
}
\decimalcolnumbers
\startdata
13    & WN10          & S       & 5.56 & 4.45  & 0.4 & 35  &   -4.69   \\
16    & WN7h          & S       & 5.80 & 4.70  & 0.3 & 42  &   -4.64   \\
22    & WN9h          & S       & 5.75 & 4.51  & 0.4 & 44  &   -4.85   \\
30    & WN6h          & S       & 5.65 & 4.67  & 0.3 & 34  &   -5.05   \\
32    & WN6(h)        & $B^{\text{x}}$   & 5.94 & 4.67  & 0.2 & 44  &   -4.63   \\
44    & WN8ha         & S       & 5.66 & 4.65  & 0.4 & 40  &   -5.12   \\
54    & WN8ha         & S       & 5.75 & 4.58  & 0.2 & 34  &   -4.97   \\
55    & WN11h         & S       & 5.77 & 4.45  & 0.4 & 45  &   -5.13   \\
58    & WN7h          & S       & 5.64 & 4.67  & 0.3 & 34  &   -5.13   \\
76    & WN9ha         & S       & 5.66 & 4.54  & 0.2 & 30  &   -5.07   \\
77    & WN7ha         & $B^{\text{c}}$  & 6.79 & 4.65  & 0.7 & 305 &   -4.87   \\
78    & WN6(+O8 V)    & $B^{?}$   & 5.70 & 4.85  & 0.2 & 32  &   -4.96   \\
79    & WN7ha+OB      & $B^{?}$   & 6.17 & 4.62  & 0.2 & 61  &   -4.46   \\
89    & WN7h          & S       & 5.78 & 4.70  & 0.2 & 35  &   -4.73   \\
91    & WN6(h)        & S       & 5.42 & 4.70  & 0.2 & 23  &   -5.15   \\
95    & WN7h+OB       & $B^{\text{x}}$   & 6.00 & 4.70  & 0.2 & 48  &   -4.21   \\
96    & WN8           & S       & 6.35 & 4.62  & 0.2 & 80  &   -4.37   \\
98    & WN6           & S       & 6.70 & 4.65  & 0.6 & 226 &   -4.43   \\
100   & WN7           & $B^{?}$   & 6.15 & 4.67  & 0.2 & 59  &   -4.52   \\
102   & WN6           & $B^{?}$   & 6.80 & 4.65  & 0.4 & 221 &   -4.21   \\
111   & WN9ha         & $B^{?}$   & 6.25 & 4.65  & 0.7 & 118 &   -5.42   \\
118   & WN6h          & $B^{\text{c}}$   & 6.66 & 4.67  & 0.2 & 136 &   -4.09   \\
119   & WN6h+?        & $B^{\text{c}}$   & 6.57 & 4.67  & 0.2 & 116 &   -4.31   \\
120   & WN9h          & S       & 5.58 & 4.51  & 0.3 & 32  &   -5.33   \\
130   & WN11h         & S       & 5.68 & 4.45  & 0.4 & 41  &   -5.35   \\
133   & WN11h         & S       & 5.69 & 4.45  & 0.4 & 41  &   -5.42   \\
\enddata 
\tablecomments{For the verification of binary (or multiple) systems, the superscripts in the third column represent: x = detected, ? = questionable, c = high X-ray emission. The catalogue of HLMC26 is taken from Table 2 and Figure 7 in the \cite{201456527Hainich}. }
\end{deluxetable*}

\section{Comparison with observations} \label{sec:observation}
\subsection{Source of the observed data} \label{sec:Source of data}
  
We filter out 18 MW-WNL stars from the Figure 2 of \citet{201962557Hamann} (hereafter HMW18), and list them in Table~\ref{tab:MW-WNL}. The data from Gaia Data Release 2 (DR2) have been corrected for distance and exhibit improved accuracy in luminosity and mass-loss rate.
Additionally, we extract 26 LMC-WNL stars from the Figure 7 of \citet{201456527Hainich} (hereafter HLMC26) and list them in Table~\ref{tab:LMC-WNL}. These samples of HLMC26 are based on the fourth catalog of WR stars in the LMC \citep{1999A&AS..137..117B}, which are also listed in Table~\ref{tab:LMC-WNL}).

\subsection{Comparison with observations} \label{sec:Comparison and analysis}

The evolutionary trajectories computed by the KO model and the observed WNL star samples are exhibited in Figure~\ref{fig:Figure(Comparison)}. We adopt the $Z = 0.02$ model (typical solar metallicity, $Z_{\odot}$) and the $Z = 0.008$ model (typical LMC metallicity, $Z_{\rm{LMC}}$) for comparison with the observed data, respectively. The distribution pattern of the observed HMW18 samples on the HR diagram generally aligns with the modeled evolutionary trends for WNL stars, though a few exceptions exist.
The effective temperature of the samples tends to be lower in the simulated HR diagram. This discrepancy arises because the $\rm{log}$ $T_{\text {eff}}$ is roughly equal to ($T_*$) measured at a stellar inner boundary radius ($R_*$) corresponding to a larger Rosseland continuum optical depth, while these observed stars have larger photospheric radii (the optical depth $\tau = 2/3$) than the model predicts. As a result, their temperatures are lower than predicted by the models \citep{2006A&A...457.1015H, 201456527Hainich, 2015A&A...581A..21H, 2023ApJS..268...51L}.

The mass range of the obversed HMW18 samples is 11-68 $\mathrm{M}_{\odot}$, which corresponds exactly to the fit of the model of $Z=0.02$ (the mass range of the modeled WNL star is 16-89 $\mathrm{M}_{\odot}$ for the KO model, and 18-85 $\mathrm{M}_{\odot}$ for the OV model, more dtails refer to the Appendix). It seems that using the KO model with a metallicity grid of 0.02 under the condition of $\eta_{\text {Dutch}} = 1.5$ can explain the distribution of WNL stars in the Milky Way more reasonably. 
But for the most LMC samples, the observed distribution is basically in the range of $\log$ $L$: 5.42-6.80, $\log$ $T_{\text{eff}}$: 4.45-4.85, and with a very high upper mass limit (e.g., according to \citet{201456527Hainich}, BAT99 77 possesses a high mass of $305$ $\mathrm{M}{\odot}$ and is likely to be identified as a binary. BAT99 98 is a single WNL star with a high mass of $226$ $\mathrm{M}{\odot}$). However, using the low metallicity ($Z = 0.008$) model to explain the distribution of the HLMC26 samples shows significant deviation. According to the previous HR diagram Figure~\ref{fig:Figure(HR)}, the model with $Z = 0.02$ and above is required to effectively account for this portion of the distribution. This implies that our current model is inadequate for explaining WNL stars in metal-poor environments.

We don't consider the influence of the fossil magnetic field of stars in this work, despite its well-known impacts on the evolution of massive stars. The magnetic field can cause mixing and diffusion of internal chemical elements, quench stellar mass loss, and induce magnetic braking \citep{2017A&A...599L...5G, 2017IAUS..329..250K, 2017MNRAS.466.1052P, 2019MNRAS.485.5843K}, there is a hypothesis that most WNL stars with low luminosity and high metallicity may be affected by magnetic fields. Additionally, it is important to note that surface rotational velocity conditions may not be universally applicable to all stars, as rotation has a significant effect on metal-poor stars. 
 
Moreover, although we precisely limited the initial metallicity and mass of the grid, it is important to acknowledge their inherent limitations in accurately reflecting actual observations. The range of initial conditions for models can introduce ambiguity, as different initial conditions may result in overlapping trajectories of stars on the HR diagram.

In addition to modeling deficiencies, these unexplained samples may also be due to observational limitations. The observed data may be related to uncertainties associated with instrument-related errors, interstellar medium extinction, contamination from neighboring stars, or tidal forces in binary systems, all of which contribute to the main sources of inaccuracy. Numerous unknown factors can significantly alter the fates of stars.
 
Recognizing the limitations between the models and the observations is essential. Therefore, the models should be viewed as a reference rather than a determinant of the initial conditions of WNL stars.

\section{Summary} \label{sec:sum}

Mixing due to overshooting and rotation, and mass loss are main uncertain factors for the massive stars. Constrained by the rapid evolutionary lifetimes of these stars and the challenges of observing their interiors, the WNL stage provides a valuable platform for studying the structure and evolution of massive stars. This is because it occurs during the early stage of the evolutionary phase and exhibits resolvable spectral characteristics.

A set of derivative equations developed by \citet{2012ApJ...756...37L, 2017ApJ...841...10L} is employed in this work to handle core overshooting, and these are compared with the classical overshooting model.
Our results provide information on the minimum initial mass required for WNL formation, the fraction of the lifetime during the WNL phase, and the mass range of the WNL stars using both overshooting schemes, and the distribution of the surface element abundances for different metallicities. This establishes a convenient data retrieval grid for future validation against a large number of observations.

The results suggest that the formation of a WNL star primarily depends on the broadening of the mixing region and the increase in the strength of the mass loss. Specifically, metal-rich massive stars with higher initial masses are more likely to evolve into WNL stars at earlier stages. Conversely, lower-mass, metal-poor single stars need to enhance stellar wind or possess surface velocities to enter the WNL phases, as rotation can facilitate the emergence of WNL stars, especially those with lower initial masses and low metallicity.

Besides the fact that the lifetimes of WNL stars increase with metallicities and initial masses, the interplay between rotation and convective overshooting also significantly influence their lifetimes and masses. 

Compared to the OV models, the KO models feature larger convective overshooting regions, which can significantly extend both the relative duration of the WNL phases and the overall lifetimes of these stars. Additionally, The KO models allow for the formation of WNL stars with higher masses. However, as the initial masses and metallicities increase, the terminal masses and the mass ranges of WNL stars obtained from the two overshooting models show little difference. The treatment of overshooting becomes the dominant factors affecting the evolutionary lifetime, while the effect of rotation becomes insignificant. This difference can be calibrated by future observations to determine which model is more accurate.

We can infer that N-rich massive stars with lower N/O ratios may originate in younger, metal-rich precursor stars. In this group, their surface N is primarily produced through the CN-cycle and partially through the ongoing NO-cycle. Conversely, stars with higher N/O ratios are likely to emerge in metal-poor environments, where their surface N primarily originates from the CN-cycle and a nearly complete NO-cycle. This suggests that these stars are likely during the late stage of the MS or the early stage of the He-MS.

Comparing the models with observations from the MW and LMC, we find that the distribution of the most observed samples aligns well with low-mass models and metallicities ranging from 0.002 to 0.02. The adoption of the KO model at $Z = 0.02$ explain the WNL samples of the MW on the HR diagram, but appears inadequate for explaining the LMC stars when using metal-poor models.
 
In conclusion, analyzing the mass, lifetime, and surface element distribution of the WNL stars using the new overshooting model enhances our understanding of how convection and overshooting influence the evolution of massive stars across different initial masses and metallicities. This study contributes significantly to unraveling the complexities of these astrophysical processes. It enables researchers to further understand the evolutionary environment and distribution characteristics of the observed samples, as well as deduce the precursor environments of the observed WNL stars. Additionally, investigating single star models with different metallicities is crucial for understanding various aspects, including the evolution of black hole progenitor stars, supernova explosions, the analysis of heavy element in the ISM, the birth of new generation stars, and the number distribution of stars in different stellar formation regions. Researchers can explore factors not considered in the models based on observations, refining them further for a more comprehensive understanding. \\

\section*{Acknowledgments}

We cordially thank the reviewer for the productive comments, which greatly helped us to improve the manuscript. We thank Dr. Xuefeng Li for his guidance on getting started with MESA code. 

This work is supported by the National Key R\&D Program of China (grant No. 2021YFA1600400/2021YFA1600402), the National Natural Science Foundation of China (Nos. 12133011, 12288102, 12273104), the Natural Science Foundation of Yunnan Province (No. 202401CF070041), the Yunnan Fundamental Research Projects (Grant No. 202401AS070045), and the International Centre of Supernovae, Yunnan Key Laboratory (No. 202302AN360001). We acknowledge the science research grants from the China Manned Space Project with No. CMS-CSST-2021-B06. The authors sincerely acknowledge the support of the Yunnan Revitalization Talent Support Program Young Talent Project.

$Software$: MESA version 12115 \citep{2011ApJS..192....3P, 2013ApJS..208....4P, 2015ApJS..220...15P, 2018ApJS..234...34P, 2019ApJS..243...10P}, PyMesaReader \citep{2017zndo....826958W}, Matplotlib \citep{2007CSE.....9...90H}, KIPPENHAHN PLOTTER \citep{2019zndo...2602098M}.

\appendix

\section{Output data by the KO and the OV models with rotation}
\vskip5pt
We present the outputs of all models with $\eta_{\mathrm{Dutch}}=1.5$ and rotation, including ages, lifetimes, and masses for the MS phase, initiation and termination of the WNL phase, initiation and termination of the WNE phase, as well as the ages and masses at the end of core-He exhaustion. Additionally, we provide the elemental abundances (Y, N, C, O) on the surface at the initiation and termination of the WNL phase. The results for the KO model and the OV model are listed in Tables~\ref{KO} and~\ref{OV}, respectively.
Note: The following definitions are applied:

- $M_{\mathrm i}$ represents the initial mass of the input model.

- $M_{\mathrm {TAMS}}$ and $t_{\mathrm {TAMS}}$ denote the terminal mass and lifetime at the termination of the MS stage. All units for the timescales in the table are expressed as ``$\mathrm{Myr}$'', signifying million years.

- $M_{\mathrm {WNL_{\mathrm i}}}$ and $t_{\mathrm {WNL_{\mathrm i}}}$ stand for the initial mass and age at the formation of the WNL star.

- $(X_{\mathrm c}/Y_{\mathrm c})_{\mathrm i}$ is the mass fraction in the core (all elemental abundances are expressed in mass fractions) indicating whether the star is in the H-burning or He-burning phase upon entering the WNL phase.

- $M_{\mathrm {WNL_{\mathrm f}}}$ and $t_{\mathrm {WNL_{\mathrm f}}}$ denote the mass and age at the termination of the WNL phase.

- $(X_{\mathrm c}/Y_{\mathrm c})_{\mathrm f}$ is the mass fraction in the core indicating whether the star is in the H-burning or He-burning phase upon leaving the WNL phase.

- $\tau_{\mathrm {WNL}}$ represents the lifetime of the WNL phase, from its beginning to the end.

- The columns 6, 9, 11-18, and 20 represent mass fractions. The surface elemental mass fractions (Y, N, C, O) at the initial and final stages of the WNL phase are marked with subscripts ``i'' and ``f'' respectively, and they are listed in columns 11-18. In columns 12-14 and 16-18, the units for surface N, C, and O mass fractions are multiplied by $10^{-4}$.

- $M_{\mathrm {WNE_{\mathrm f}}}$ represents the mass at the termination of the WNE phase.

- ${Y_{\mathrm c}}_{\mathrm {WNE_{\mathrm f}}}$ indicates the corresponding He mass fraction (expressed in percentage terms) during the He-burning phase at the termination of the WNE phase.

- $\tau_{\mathrm {WNE}}$ represents the lifetime of the WNE phase.

- $M_{\mathrm {He_{\mathrm f}}}$ and $t_{\mathrm {He_f}}$ denote the final mass and age at the termination of the He-burning stage.

- $\frac{\tau_{\mathrm {WNL}}}{t_{\mathrm {WNL_{\mathrm f}}}}$ represents the proportion of the WNL phase's lifetime to the total lifetime at the termination of the WNL phase.

\vskip12pt

\begin{longrotatetable}
\begin{deluxetable*}{cccccccccccccccccccccccc}
\tablecaption{Output data by the KO with rotation \label{KO}}
%\tablewidth{700pt}
\setlength{\tabcolsep}{2pt} 
\tabletypesize{\scriptsize}
\tablehead{\colhead{$M_{\mathrm i}$} &    \colhead{$M_{\mathrm {TAMS}}$} &    \colhead{$t_{\mathrm {TAMS}}$}  &    \colhead{$M_{\mathrm {WNL_{\mathrm i}}}$} &    \colhead{$t_{\mathrm {WNL_{\mathrm i}}}$}  &     \colhead{$(X/Y)_{\mathrm i}$}   &     \colhead{$M_{\mathrm {WNL_{\mathrm f}}}$}  &     \colhead{$t_{\mathrm {WNL_{\mathrm f}}}$} &    \colhead{$(X/Y)_{\mathrm f}$}  &     \colhead{$\tau_{\mathrm {WNL}}$}   & \colhead{$Y_{s_{\mathrm i}}$} &    \colhead{N$_{s_{\mathrm i}}$} &   \colhead{C$_{s_{\mathrm i}}$}  &   \colhead{O$_{s_{\mathrm i}}$} &   \colhead{$Y_{s_{\mathrm f}}$}  &    \colhead{N$_{s_{\mathrm f}}$} &   \colhead{C$_{s_{\mathrm f}}$}  &   \colhead{O$_{s_{\mathrm f}}$} &    \colhead{$M_{\mathrm {WNE_{\mathrm f}}}$}  &     \colhead{$Y_{\mathrm {WNE_{\mathrm f}}}$} &    \colhead{$\tau_{\mathrm {WNE}}$}  &     \colhead{$M_{\mathrm {He_{\mathrm f}}}$} &   \colhead{$t_{\mathrm {He_f}}$}   &     \colhead{$\frac{\tau_{\mathrm {WNL}}}{t_{\mathrm {WNL_{\mathrm f}}}}$} \\ 
\colhead{$(\mathrm{M}_{\odot})$}  &    \colhead{$(\mathrm{M}_{\odot})$}  &    \colhead{$(Myr)$}  &    \colhead{$(\mathrm{M}_{\odot})$}  &    \colhead{$(Myr)$}  &     \colhead{ }  &  \colhead{$(\mathrm{M}_{\odot})$}  & 
    \colhead{$(Myr)$}  &   \colhead{ }  &  \colhead{$(Myr)$ } & \colhead{ }  &    \multicolumn{3}{c}{(mass fract.* ${10^{-4}}$) }   &  \colhead{ }  &  \multicolumn{3}{c}{(mass fract.* ${10^{-4}}$)}   &     \colhead{$(\mathrm{M}_{\odot})$}  &   \colhead{ }  &     \colhead{$(Myr)$}  &     \colhead{$(\mathrm{M}_{\odot})$}  &     \colhead{$(Myr)$}   &    \colhead{(\%)}   
}  
%\decimalcolnumbers
\startdata 
\multicolumn{24}{c}{$Z = 0.002$}\\
\hline
50   & 46  & 4.58 & \nodata & \nodata & \nodata & \nodata & \nodata & \nodata & \nodata & \nodata & \nodata & \nodata & \nodata & \nodata & \nodata & \nodata &  \nodata & \nodata & \nodata & \nodata & 38 & 4.97& \nodata  \\
60   & 53  & 4.22&  \nodata & \nodata & \nodata & \nodata & \nodata & \nodata & \nodata & \nodata & \nodata & \nodata & \nodata & \nodata & \nodata & \nodata & \nodata & \nodata & \nodata & \nodata & 43 & 4.60&   \nodata  \\
70   & 61  & 3.88&   49 & 3.96  & 0.70(Y)& 37 & 4.21  & 0.04(Y)& 0.25& 0.60  & 12.50    & 0.27     & 0.46     & 0.94 & 12.79      & 0.18       & 0.25       & 36 & 0.00& 0.02& 36 & 4.23&   5.84     \\
80   & 69  & 3.64&   57 & 3.82  & 0.33(Y)& 48 & 3.97  & 0.00(Y)& 0.15& 0.60  & 12.26    & 0.35     & 0.62     & 0.79 & 12.69      & 0.27       & 0.24       & \nodata   &  \nodata   & 0.00& 48 & 3.97&   3.80     \\
90   & 76  & 3.45&   62 & 3.76  & 0.01(Y)& 62 & 3.77  & 0.00(Y)& 0.01& 0.60  & 12.32    & 0.34     & 0.57     & 0.68 & 12.49      & 0.30       & 0.42       &  \nodata  &  \nodata   & 0.00& 62 & 3.77&   0.15     \\
100  & 84  & 3.30&   67 & 3.59  & 0.02(Y)& 66 & 3.61  & 0.00(Y)& 0.01& 0.60  & 12.41    & 0.31     & 0.51     & 0.68 & 12.53      & 0.28       & 0.40       &  \nodata  &  \nodata   & 0.00& 66 & 3.61&   0.33     \\
110  & 91  & 3.17&   80 & 3.44  & 0.07(Y)& 76 & 3.48  & 0.00(Y)& 0.04& 0.60  & 12.44    & 0.31     & 0.46     & 0.79 & 12.68      & 0.28       & 0.24       &  \nodata  &   \nodata  & 0.00& 76 & 3.48&   1.07     \\
120  & 100 & 3.04&   82 & 3.33  & 0.01(Y)& 81 & 3.33  & 0.00(Y)& 0.00& 0.60  & 12.58    & 0.28     & 0.36     & 0.60 & 12.58      & 0.28       & 0.36       &  \nodata  &  \nodata   & 0.00& 81 & 3.33&   0.09     \\
130  & 106 & 2.98&   102& 3.03  & 0.77(Y)& 77 & 3.19  & 0.19(Y)& 0.16& 0.60  & 12.57    & 0.28     & 0.36     & 0.94 & 12.79      & 0.19       & 0.23       & 73 & 0.13& 0.02& 63 & 3.27&   4.96     \\
140  & 114 & 2.90&   113& 2.91  & 0.95(Y)& 83 & 3.09  & 0.22(Y)& 0.18& 0.61  & 12.62    & 0.27     & 0.32     & 0.95 & 12.79      & 0.19       & 0.23       & 78 & 0.16& 0.02& 65 & 3.18&   5.72     \\
150  & 120 & 2.83&   120& 2.84  & 0.99(Y)& 90 & 2.99  & 0.32(Y)& 0.15& 0.60  & 12.66    & 0.26     & 0.28     & 0.95 & 12.79      & 0.19       & 0.22       & 85 & 0.26& 0.02& 59 & 3.12&   5.11     \\
\hline
\multicolumn{24}{c}{$Z = 0.008$}\\
\hline
50   & 39  & 4.48&   28 & 4.58  & 0.70(Y)& 23 & 4.71  & 0.33(Y)& 0.13& 0.61  & 50.34    & 0.83     & 1.77     & 0.93 & 51.13      & 0.68       & 1.07       & 20 & 0.18& 0.06& 17 & 4.88&   2.79     \\
60   & 44  & 4.10&   35 & 4.19  & 0.71(Y)& 28 & 4.30  & 0.38(Y)& 0.11& 0.59  & 50.02    & 0.94     & 1.98     & 0.94 & 51.13      & 0.71       & 1.03       & 25 & 0.26& 0.04& 19 & 4.48&   2.61     \\
70   & 51  & 3.78&   44 & 3.89  & 0.64(Y)& 34 & 4.01  & 0.28(Y)& 0.12& 0.59  & 50.57    & 0.86     & 1.46     & 0.94 & 51.12      & 0.74       & 1.01       & 31 & 0.20& 0.03& 25 & 4.14&   3.01     \\
80   & 54  & 3.55&   52 & 3.59  & 0.88(Y)& 40 & 3.71  & 0.44(Y)& 0.12& 0.59  & 50.74    & 0.85     & 1.29     & 0.93 & 51.11      & 0.74       & 1.01       & 36 & 0.35& 0.03& 23 & 3.90&   3.25     \\
90   & 56  & 3.38&   57 & 3.36  & 0.01(X)& 43 & 3.45  & 0.71(Y)& 0.10& 0.60  & 50.75    & 0.86     & 1.26     & 0.94 & 51.11      & 0.73       & 1.01       & 40 & 0.60& 0.03& 15 & 3.75&   2.81     \\
100  & 45  & 3.26&   65 & 3.06  & 0.07(X)& 48 & 3.23  & 0.01(X)& 0.17& 0.59  & 50.75    & 0.87     & 1.25     & 0.93 & 50.76      & 1.03       & 1.02       & 39 & 0.83& 0.06& 10 & 3.68&   5.27     \\
110  & 37  & 3.16&   72 & 2.87  & 0.10(X)& 51 & 3.05  & 0.04(X)& 0.18& 0.60  & 50.74    & 0.88     & 1.23     & 0.94 & 50.76      & 1.05       & 1.00       & 31 & 0.77& 0.17& 10 & 3.59&   5.96     \\
120  & 34  & 3.08&   79 & 2.71  & 0.13(X)& 45 & 2.97  & 0.04(X)& 0.26& 0.60  & 50.74    & 0.89     & 1.23     & 0.94 & 50.76      & 1.04       & 1.01       & 29 & 0.74& 0.19& 10 & 3.51&   8.79     \\
130  & 33  & 3.01&   86 & 2.58  & 0.14(X)& 42 & 2.90  & 0.04(X)& 0.32& 0.60  & 50.75    & 0.90     & 1.21     & 0.94 & 50.77      & 1.03       & 1.01       & 27 & 0.73& 0.19& 10 & 3.44&   10.93    \\
140  & 30  & 2.96&   95 & 2.42  & 0.17(X)& 37 & 2.86  & 0.03(X)& 0.44& 0.60  & 50.75    & 0.90     & 1.20     & 0.94 & 50.77      & 1.02       & 1.02       & 25 & 0.70& 0.19& 10 & 3.40&   15.48    \\
150  & 30  & 2.91&   102& 2.33  & 0.19(X)& 37 & 2.80  & 0.03(X)& 0.48& 0.59  & 50.75    & 0.91     & 1.19     & 0.94 & 50.77      & 1.02       & 1.02       & 25 & 0.69& 0.20& 10 & 3.35&   16.97    \\
\hline
\multicolumn{24}{c}{$Z = 0.014$}\\
\hline
50   & 34  & 4.43&   26 & 4.52  & 0.72(Y)& 22 & 4.59  & 0.54(Y)& 0.07& 0.59  & 87.16    & 1.31     & 4.35     & 0.93 & 89.37      & 1.21       & 1.96       & 19 & 0.36& 0.06& 12 & 4.86&   1.52     \\
60   & 42  & 4.00&   36 & 4.08  & 0.76(Y)& 26 & 4.21  & 0.36(Y)& 0.13& 0.59  & 86.81    & 1.44     & 4.57     & 0.93 & 89.37      & 1.26       & 1.90       & 23 & 0.25& 0.04& 18 & 4.39&   3.19     \\
70   & 42  & 3.72&   42 & 3.73  & 0.98(Y)& 32 & 3.82  & 0.66(Y)& 0.09& 0.59  & 88.38    & 1.39     & 2.85     & 0.93 & 89.32      & 1.27       & 1.93       & 28 & 0.53& 0.04& 13 & 4.11&   2.48     \\
80   & 30  & 3.53&   48 & 3.27  & 0.08(X)& 34 & 3.49  & 0.01(X)& 0.21& 0.59  & 88.45    & 1.42     & 2.73     & 0.93 & 88.76      & 1.71       & 1.99       & 24 & 0.76& 0.11& 8  & 3.99&   6.07     \\
90   & 24  & 3.40&   54 & 2.94  & 0.13(X)& 30 & 3.29  & 0.03(X)& 0.35& 0.59  & 88.45    & 1.44     & 2.69     & 0.94 & 88.79      & 1.69       & 1.98       & 18 & 0.67& 0.22& 8  & 3.89&   10.52    \\
100  & 22  & 3.32&   62 & 2.70  & 0.17(X)& 25 & 3.24  & 0.02(X)& 0.55& 0.59  & 88.48    & 1.46     & 2.64     & 0.94 & 88.80      & 1.67       & 2.00       & 17 & 0.63& 0.21& 8  & 3.81&   16.89    \\
110  & 21  & 3.23&   67 & 2.56  & 0.18(X)& 25 & 3.14  & 0.02(X)& 0.58& 0.59  & 88.50    & 1.48     & 2.59     & 0.93 & 88.81      & 1.66       & 2.00       & 16 & 0.62& 0.22& 8  & 3.73&   18.43    \\
120  & 21  & 3.17&   75 & 2.38  & 0.21(X)& 23 & 3.10  & 0.02(X)& 0.73& 0.59  & 88.50    & 1.48     & 2.59     & 0.94 & 88.80      & 1.66       & 2.01       & 16 & 0.60& 0.21& 8  & 3.67&   23.40    \\
130  & 20  & 3.12&   82 & 2.24  & 0.23(X)& 22 & 3.06  & 0.02(X)& 0.82& 0.59  & 88.51    & 1.49     & 2.56     & 0.94 & 88.80      & 1.65       & 2.02       & 15 & 0.58& 0.21& 8  & 3.62&   26.91    \\
140  & 20  & 3.07&   88 & 2.14  & 0.25(X)& 22 & 3.01  & 0.01(X)& 0.87& 0.59  & 88.54    & 1.50     & 2.51     & 0.94 & 88.80      & 1.65       & 2.02       & 15 & 0.57& 0.21& 8  & 3.57&   29.01    \\
150  & 20  & 3.03&   95 & 2.03  & 0.26(X)& 21 & 2.97  & 0.01(X)& 0.94& 0.59  & 88.57    & 1.51     & 2.47     & 0.94 & 88.80      & 1.65       & 2.03       & 15 & 0.56& 0.21& 8  & 3.53&   31.65    \\
\hline
\multicolumn{24}{c}{$Z = 0.02$}\\
\hline
50   & 32  & 4.27&   26 & 4.34  & 0.78(Y)& 21 & 4.41  & 0.61(Y)& 0.06& 0.58  & 122.15   & 1.75     & 9.08     & 0.93 & 127.41     & 1.76       & 3.05       & 18 & 0.43& 0.06& 10 & 4.71&   1.45     \\
60   & 38  & 3.88&   33 & 3.94  & 0.83(Y)& 26 & 4.01  & 0.59(Y)& 0.08& 0.58  & 123.34   & 1.85     & 7.58     & 0.93 & 127.41     & 1.82       & 2.97       & 22 & 0.45& 0.04& 12 & 4.30&   1.94     \\
70   & 27  & 3.65&   40 & 3.41  & 0.07(X)& 28 & 3.64  & 0.00(X)& 0.23& 0.59  & 124.53   & 1.92     & 6.14     & 0.93 & 126.68     & 2.35       & 3.11       & 21 & 0.76& 0.09& 7  & 4.14&   6.32     \\
80   & 19  & 3.52&   46 & 2.99  & 0.14(X)& 23 & 3.43  & 0.02(X)& 0.44& 0.58  & 124.63   & 1.96     & 5.97     & 0.93 & 126.74     & 2.30       & 3.10       & 14 & 0.62& 0.23& 7  & 4.06&   12.94    \\
90   & 17  & 3.44&   53 & 2.68  & 0.19(X)& 19 & 3.39  & 0.01(X)& 0.71& 0.58  & 124.64   & 1.98     & 5.93     & 0.93 & 126.76     & 2.26       & 3.13       & 12 & 0.56& 0.22& 6  & 3.99&   20.99    \\
100  & 17  & 3.36&   58 & 2.49  & 0.21(X)& 18 & 3.33  & 0.01(X)& 0.84& 0.59  & 124.80   & 2.00     & 5.72     & 0.93 & 126.76     & 2.25       & 3.14       & 12 & 0.54& 0.22& 6  & 3.92&   25.29    \\
110  & 16  & 3.30&   65 & 2.29  & 0.24(X)& 17 & 3.28  & 0.00(X)& 0.99& 0.58  & 124.80   & 2.02     & 5.70     & 0.93 & 126.77     & 2.24       & 3.15       & 11 & 0.52& 0.22& 6  & 3.87&   30.11    \\
120  & 16  & 3.24&   71 & 2.17  & 0.25(X)& 16 & 3.22  & 0.00(X)& 1.06& 0.59  & 124.95   & 2.04     & 5.50     & 0.93 & 126.77     & 2.24       & 3.15       & 11 & 0.51& 0.22& 6  & 3.81&   32.83    \\
130  & 16  & 3.19&   78 & 2.04  & 0.27(X)& 16 & 3.19  & 0.00(X)& 1.15& 0.59  & 125.01   & 2.05     & 5.41     & 0.93 & 126.77     & 2.23       & 3.16       & 11 & 0.50& 0.22& 6  & 3.76&   36.15    \\
140  & 16  & 3.15&   84 & 1.94  & 0.28(X)& 16 & 3.14  & 0.00(X)& 1.20& 0.58  & 124.98   & 2.06     & 5.43     & 0.93 & 126.77     & 2.23       & 3.16       & 11 & 0.50& 0.21& 6  & 3.72&   38.33    \\
150  & 16  & 3.10&   89 & 1.87  & 0.29(X)& 16 & 3.09  & 0.00(X)& 1.21& 0.59  & 125.11   & 2.08     & 5.26     & 0.93 & 126.77     & 2.23       & 3.15       & 11 & 0.50& 0.22& 6  & 3.67&   39.33    \\
\hline
\multicolumn{24}{c}{$Z = 0.03$}\\
\hline
50   & 32  & 4.06&   25 & 4.14  & 0.77(Y)& 21 & 4.20  & 0.61(Y)& 0.06& 0.57  & 173.95   & 2.46     & 24.44    & 0.91 & 189.62     & 2.67       & 6.25       & 18 & 0.46& 0.05& 9  & 4.52&   1.34     \\
60   & 27  & 3.74&   33 & 3.59  & 0.04(X)& 23 & 3.78  & 0.89(Y)& 0.19& 0.57  & 177.86   & 2.62     & 19.77    & 0.92 & 189.63     & 2.68       & 6.23       & 19 & 0.71& 0.06& 7  & 4.26&   4.99     \\
70   & 16  & 3.62&   39 & 2.99  & 0.15(X)& 17 & 3.57  & 0.01(X)& 0.59& 0.58  & 178.32   & 2.68     & 19.15    & 0.92 & 189.04     & 3.25       & 6.14       & 10 & 0.60& 0.22& 5  & 4.25&   16.42    \\
80   & 14  & 3.54&   45 & 2.61  & 0.20(X)& 13 & 3.55  & 0.97(Y)& 0.94& 0.57  & 178.42   & 2.72     & 18.99    & 0.92 & 189.11     & 3.19       & 6.15       & 9  & 0.53& 0.21& 5  & 4.21&   26.44    \\
90   & 13  & 3.48&   50 & 2.35  & 0.24(X)& 12 & 3.50  & 0.96(Y)& 1.15& 0.57  & 178.79   & 2.75     & 18.52    & 0.92 & 189.14     & 3.16       & 6.15       & 8  & 0.50& 0.22& 5  & 4.15&   32.77    \\
100  & 13  & 3.42&   56 & 2.16  & 0.26(X)& 12 & 3.44  & 0.96(Y)& 1.28& 0.57  & 178.87   & 2.78     & 18.39    & 0.92 & 189.18     & 3.14       & 6.13       & 8  & 0.49& 0.23& 5  & 4.10&   37.19    \\
110  & 13  & 3.36&   62 & 2.01  & 0.28(X)& 12 & 3.39  & 0.95(Y)& 1.38& 0.57  & 179.06   & 2.81     & 18.14    & 0.92 & 189.21     & 3.12       & 6.12       & 8  & 0.48& 0.23& 5  & 4.05&   40.70    \\
120  & 13  & 3.31&   67 & 1.90  & 0.29(X)& 12 & 3.33  & 0.95(Y)& 1.43& 0.58  & 179.71   & 2.84     & 17.36    & 0.92 & 189.25     & 3.12       & 6.08       & 8  & 0.48& 0.23& 5  & 3.99&   42.85    \\
130  & 13  & 3.27&   73 & 1.78  & 0.31(X)& 12 & 3.29  & 0.95(Y)& 1.51& 0.58  & 179.88   & 2.86     & 17.13    & 0.92 & 189.27     & 3.10       & 6.07       & 8  & 0.48& 0.23& 7  & 3.57&   45.88    \\
140  & 13  & 3.22&   78 & 1.71  & 0.31(X)& 12 & 3.25  & 0.95(Y)& 1.54& 0.57  & 179.87   & 2.87     & 17.13    & 0.92 & 189.28     & 3.12       & 6.05       & 8  & 0.48& 0.23& 8  & 3.51&   47.46    \\
150  & 13  & 3.19&   84 & 1.61  & 0.33(X)& 12 & 3.22  & 0.95(Y)& 1.60& 0.57  & 179.83   & 2.89     & 17.15    & 0.92 & 189.33     & 3.08       & 6.03       & 8  & 0.48& 0.23& 5  & 3.74&   49.87    \\
\hline
\multicolumn{24}{c}{$Z = 0.04$}\\
\hline
50   & 31  & 3.86&   25 & 3.93  & 0.78(Y)& 20 & 3.99  & 0.62(Y)& 0.06& 0.58  & 220.53   & 3.13     & 45.81    & 0.88 & 247.78     & 3.47       & 14.23      & 17 & 0.49& 0.05& 8  & 4.33&   1.47     \\
60   & 15  & 3.67&   33 & 3.12  & 0.13(X)& 17 & 3.60  & 0.02(X)& 0.48& 0.56  & 221.88   & 3.22     & 44.16    & 0.91 & 248.24     & 4.18       & 12.75      & 10 & 0.65& 0.22& 4  & 4.36&   13.36    \\
70   & 13  & 3.59&   38 & 2.68  & 0.19(X)& 12 & 3.61  & 0.96(Y)& 0.93& 0.57  & 223.44   & 3.30     & 42.27    & 0.91 & 248.56     & 4.06       & 12.55      & 8  & 0.55& 0.21& 4  & 4.32&   25.85    \\
80   & 12  & 3.54&   44 & 2.32  & 0.23(X)& 10 & 3.58  & 0.93(Y)& 1.26& 0.56  & 223.78   & 3.35     & 41.82    & 0.91 & 248.96     & 3.74       & 12.52      & 7  & 0.49& 0.24& 4  & 4.29&   35.08    \\
90   & 11  & 3.48&   49 & 2.08  & 0.26(X)& 10 & 3.53  & 0.91(Y)& 1.45& 0.56  & 224.12   & 3.39     & 41.38    & 0.91 & 249.71     & 3.14       & 12.46      & 7  & 0.47& 0.24& 4  & 4.24&   41.06    \\
100  & 11  & 3.43&   54 & 1.92  & 0.28(X)& 10 & 3.48  & 0.91(Y)& 1.56& 0.57  & 224.80   & 3.43     & 40.55    & 0.91 & 249.78     & 3.13       & 12.39      & 7  & 0.46& 0.25& 4  & 4.18&   44.81    \\
110  & 11  & 3.39&   60 & 1.76  & 0.30(X)& 10 & 3.45  & 0.90(Y)& 1.69& 0.56  & 224.92   & 3.45     & 40.37    & 0.91 & 249.86     & 3.10       & 12.34      & 7  & 0.45& 0.25& 4  & 4.15&   49.03    \\
120  & 11  & 3.32&   65 & 1.67  & 0.31(X)& 10 & 3.38  & 0.90(Y)& 1.71& 0.56  & 225.23   & 3.48     & 39.98    & 0.91 & 249.94     & 3.11       & 12.23      & 7  & 0.45& 0.25& 4  & 4.08&   50.53    \\
130  & 11  & 3.29&   71 & 1.56  & 0.33(X)& 10 & 3.35  & 0.90(Y)& 1.79& 0.56  & 225.68   & 3.51     & 39.42    & 0.91 & 249.99     & 3.10       & 12.19      & 7  & 0.45& 0.25& 4  & 4.05&   53.33    \\
140  & 11  & 3.24&   75 & 1.51  & 0.33(X)& 10 & 3.30  & 0.90(Y)& 1.79& 0.56  & 225.76   & 3.53     & 39.31    & 0.91 & 250.05     & 3.11       & 12.11      & 7  & 0.45& 0.25& 4  & 4.00&   54.37    \\
150  & 11  & 3.22&   81 & 1.41  & 0.34(X)& 10 & 3.28  & 0.90(Y)& 1.87& 0.56  & 225.84   & 3.55     & 39.19    & 0.91 & 250.09     & 3.11       & 12.07      & 7  & 0.45& 0.25& 4  & 3.99&   56.93    \\
\enddata
\end{deluxetable*}
\end{longrotatetable}

\begin{longrotatetable}
\begin{deluxetable*}{cccccccccccccccccccccccc}
\tablecaption{Output data by the OV with rotation \label{OV}}
%\tablewidth{700pt}
\setlength{\tabcolsep}{2pt} 
\tabletypesize{\scriptsize}
\tablehead{\colhead{$M_{\mathrm i}$} &    \colhead{$M_{\mathrm {TAMS}}$} &    \colhead{$t_{\mathrm {TAMS}}$}  &    \colhead{$M_{\mathrm {WNL_{\mathrm i}}}$} &    \colhead{$t_{\mathrm {WNL_{\mathrm i}}}$}  &     \colhead{$(X/Y)_{\mathrm i}$}   &     \colhead{$M_{\mathrm {WNL_{\mathrm f}}}$}  &     \colhead{$t_{\mathrm {WNL_{\mathrm f}}}$} &    \colhead{$(X/Y)_{\mathrm f}$}  &     \colhead{$\tau_{\mathrm {WNL}}$}   & \colhead{$Y_{s_{\mathrm i}}$} &    \colhead{N$_{s_{\mathrm i}}$} &   \colhead{C$_{s_{\mathrm i}}$}  &   \colhead{O$_{s_{\mathrm i}}$} &   \colhead{$Y_{s_{\mathrm f}}$}  &    \colhead{N$_{s_{\mathrm f}}$} &   \colhead{C$_{s_{\mathrm f}}$}  &   \colhead{O$_{s_{\mathrm f}}$} &    \colhead{$M_{\mathrm {WNE_{\mathrm f}}}$}  &     \colhead{$Y_{\mathrm {WNE_{\mathrm f}}}$} &    \colhead{$\tau_{\mathrm {WNE}}$}  &     \colhead{$M_{\mathrm {He_{\mathrm f}}}$} &   \colhead{$t_{\mathrm {He_f}}$}   &     \colhead{$\frac{\tau_{\mathrm {WNL}}}{t_{\mathrm {WNL_{\mathrm f}}}}$} \\ 
\colhead{$(\mathrm{M}_{\odot})$}  &    \colhead{$(\mathrm{M}_{\odot})$}  &    \colhead{$(Myr)$}  &    \colhead{$(\mathrm{M}_{\odot})$}  &    \colhead{$(Myr)$}  &     \colhead{ }  &  \colhead{$(\mathrm{M}_{\odot})$}  & 
    \colhead{$(Myr)$}  &   \colhead{ }  &  \colhead{$(Myr)$ } & \colhead{ }  &    \multicolumn{3}{c}{(mass fract.* ${10^{-4}}$) }   &  \colhead{ }  &  \multicolumn{3}{c}{(mass fract.* ${10^{-4}}$)}   &     \colhead{$(\mathrm{M}_{\odot})$}  &   \colhead{ }  &     \colhead{$(Myr)$}  &     \colhead{$(\mathrm{M}_{\odot})$}  &     \colhead{$(Myr)$}   &    \colhead{(\%)}   
} 
%\decimalcolnumbers
\startdata 
\multicolumn{24}{c}{$Z = 0.002$}\\
\hline
50   & 45  & 4.65 & \nodata & \nodata & \nodata & \nodata & \nodata & \nodata & \nodata & \nodata & \nodata & \nodata & \nodata & \nodata & \nodata & \nodata & \nodata & \nodata & \nodata & \nodata & 39 & 5.03&   \nodata  \\
60   & 53  & 4.16&   41 & 4.43  & 0.16(Y)& 39 & 4.52  & 0.00(Y)& 0.09& 0.60  & 12.06    & 0.38     & 0.81     & 0.65 & 12.35      & 0.31       & 0.57      & \nodata & \nodata & \nodata & 39 & 4.52&   2.02     \\
70   & 61  & 3.83&   48 & 4.11  & 0.11(Y)& 46 & 4.17  & 0.00(Y)& 0.06& 0.60  & 12.72    & 0.21     & 0.28     & 0.71 & 12.68      & 0.26       & 0.26       & \nodata & \nodata & \nodata & 46 & 4.17&   1.48     \\
80   & 70  & 3.54&   54 & 3.75  & 0.24(Y)& 49 & 3.86  & 0.00(Y)& 0.11& 0.60  & 12.57    & 0.25     & 0.40     & 0.73 & 12.67      & 0.27       & 0.26       & \nodata & \nodata & \nodata & 49 & 3.86&   2.98     \\
90   & 77  & 3.37&   64 & 3.58  & 0.23(Y)& 57 & 3.69  & 0.00(Y)& 0.11& 0.60  & 12.72    & 0.22     & 0.27     & 0.75 & 12.68      & 0.28       & 0.24       & \nodata & \nodata & \nodata & 57 & 3.69&   2.95     \\
100  & 85  & 3.20&   71 & 3.46  & 0.08(Y)& 68 & 3.50  & 0.00(Y)& 0.04& 0.60  & 12.61    & 0.26     & 0.34     & 0.68 & 12.68      & 0.26       & 0.26       & \nodata & \nodata & \nodata & 68 & 3.50&   1.16     \\
110  & 93  & 3.08&   78 & 3.35  & 0.06(Y)& 75 & 3.37  & 0.00(Y)& 0.03& 0.60  & 12.61    & 0.26     & 0.35     & 0.68 & 12.68      & 0.26       & 0.27       & \nodata & \nodata & \nodata & 75 & 3.37&   0.83     \\
120  & 101 & 2.99&   83 & 3.24  & 0.07(Y)& 80 & 3.28  & 0.00(Y)& 0.04 & 0.60  & 12.50    & 0.29     & 0.43     & 0.68 & 12.65  & 0.26       & 0.29       & \nodata & \nodata & \nodata & 80 & 3.28&   1.10     \\
130  & 108 & 2.91&  \nodata  &  \nodata   & \nodata  & \nodata  &  \nodata  & \nodata  & \nodata   & \nodata  & \nodata  & \nodata    & \nodata  & \nodata  &    \nodata & \nodata  &    \nodata  & \nodata  & \nodata  & \nodata & 92 & 3.19&     \nodata    \\
140  & 116 & 2.84&   102& 3.11  & 0.01(Y)& 101& 3.12  & 0.00(Y)& 0.01& 0.60  & 12.56    & 0.29     & 0.36     & 0.73 & 12.67      & 0.28       & 0.26       & \nodata & \nodata & \nodata & 101& 3.11&   0.18     \\
150  & 123 & 2.78&   111& 2.95  & 0.27(Y)& 95 & 3.05  & 0.00(Y)& 0.11& 0.60  & 12.60    & 0.27     & 0.33     & 0.81 & 12.71      & 0.27       & 0.22       & \nodata & \nodata & \nodata &  95 & 3.05&   3.48     \\
\hline
\multicolumn{24}{c}{$Z = 0.008$}\\
\hline
50   & 41  & 4.45&   29 & 4.61  & 0.51(Y)& 22 & 4.85  & 0.02(Y)& 0.24& 0.61  & 50.66    & 0.77     & 1.48     & 0.93 & 51.07      & 0.71       & 1.10       & 21 & 0.00& 0.01& 21 & 4.86&   4.90     \\
60   & 46  & 3.98&   34 & 4.09  & 0.64(Y)& 27 & 4.25  & 0.20(Y)& 0.16& 0.63  & 50.71    & 0.82     & 1.37     & 0.91 & 51.09      & 0.70       & 1.08       & 23 & 0.09& 0.05& 22 & 4.36&   3.69     \\
70   & 52  & 3.70&   43 & 3.82  & 0.59(Y)& 32 & 3.98  & 0.14(Y)& 0.16& 0.60  & 50.58    & 0.85     & 1.46     & 0.92 & 51.10      & 0.71       & 1.06       & 28 & 0.06& 0.04& 28 & 4.05&   4.13     \\
80   & 57  & 3.45&   48 & 3.56  & 0.58(Y)& 37 & 3.68  & 0.24(Y)& 0.11& 0.59  & 50.69    & 0.85     & 1.34     & 0.92 & 51.11      & 0.73       & 1.03       & 33 & 0.15& 0.03& 28 & 3.79&   3.04     \\
90   & 61  & 3.29&   55 & 3.37  & 0.69(Y)& 42 & 3.48  & 0.32(Y)& 0.11& 0.59  & 50.75    & 0.85     & 1.27     & 0.94 & 51.12      & 0.74       & 0.99       & 37 & 0.23& 0.03& 29 & 3.62&   3.15     \\
100  & 64  & 3.15&   63 & 3.16  & 0.96(Y)& 46 & 3.28  & 0.50(Y)& 0.11& 0.59  & 50.76    & 0.86     & 1.25     & 0.94 & 51.13      & 0.75       & 0.98       & 42 & 0.39& 0.03& 24 & 3.48&   3.50     \\
110  & 65  & 3.04&   69 & 3.00  & 0.02(X)& 51 & 3.12  & 0.70(Y)& 0.12& 0.59  & 50.76    & 0.87     & 1.23     & 0.94 & 51.14      & 0.75       & 0.97       & 45 & 0.58& 0.03& 17 & 3.40&   3.77     \\
120  & 61  & 2.95&   75 & 2.82  & 0.05(X)& 55 & 2.98  & 0.92(Y)& 0.15& 0.59  & 50.76    & 0.88     & 1.22     & 0.94 & 51.10      & 0.75       & 1.01       & 48 & 0.76& 0.04& 13 & 3.34&   5.07     \\
130  & 50  & 2.89&   82 & 2.68  & 0.08(X)& 60 & 2.84  & 0.02(X)& 0.16& 0.60  & 50.76    & 0.89     & 1.21     & 0.94 & 50.76      & 1.05       & 1.00       & 43 & 0.80& 0.09& 11 & 3.29&   5.69     \\
140  & 44  & 2.83&   89 & 2.57  & 0.10(X)& 64 & 2.73  & 0.04(X)& 0.16& 0.60  & 50.76    & 0.90     & 1.20     & 0.93 & 50.76      & 1.05       & 1.00       & 37 & 0.77& 0.16& 11 & 3.24&   5.75     \\
150  & 39  & 2.79&   97 & 2.45  & 0.12(X)& 56 & 2.67  & 0.04(X)& 0.22& 0.59  & 50.76    & 0.91     & 1.19     & 0.94 & 50.76      & 1.05       & 0.99       & 33 & 0.73& 0.19& 11 & 3.20&   8.24     \\
\hline
\multicolumn{24}{c}{$Z = 0.014$}\\
\hline
50   & 37  & 4.31&   28 & 4.41  & 0.72(Y)& 21 & 4.56  & 0.32(Y)& 0.15& 0.63  & 87.97    & 1.30     & 3.44     & 0.84 & 89.17      & 1.18       & 2.23       & 18 & 0.15& 0.07& 15 & 4.73&   3.24     \\
60   & 43  & 3.91&   35 & 4.00  & 0.69(Y)& 26 & 4.17  & 0.22(Y)& 0.17& 0.59  & 87.38    & 1.37     & 4.02     & 0.79 & 89.07      & 1.17       & 2.36       & 22 & 0.12& 0.04& 20 & 4.29&   4.03     \\
70   & 45  & 3.63&   39 & 3.69  & 0.76(Y)& 30 & 3.78  & 0.49(Y)& 0.08& 0.59  & 88.06    & 1.36     & 3.26     & 0.89 & 89.25      & 1.27       & 2.01       & 26 & 0.35& 0.04& 17 & 4.00&   2.19     \\
80   & 47  & 3.40&   47 & 3.41  & 0.98(Y)& 34 & 3.52  & 0.58(Y)& 0.11& 0.59  & 88.44    & 1.40     & 2.77     & 0.93 & 89.37      & 1.29       & 1.85       & 30 & 0.45& 0.04& 16 & 3.77&   3.02     \\
90   & 37  & 3.27&   52 & 3.08  & 0.07(X)& 36 & 3.27  & 0.99(Y)& 0.19& 0.60  & 88.48    & 1.44     & 2.67     & 0.93 & 88.77      & 1.72       & 1.97       & 29 & 0.75& 0.07& 10 & 3.70&   5.90     \\
100  & 30  & 3.15&   58 & 2.83  & 0.10(X)& 39 & 3.05  & 0.03(X)& 0.22& 0.59  & 88.51    & 1.45     & 2.62     & 0.93 & 88.78      & 1.73       & 1.95       & 23 & 0.70& 0.18& 9  & 3.60&   7.20     \\
110  & 26  & 3.07&   64 & 2.66  & 0.13(X)& 34 & 2.96  & 0.03(X)& 0.30& 0.59  & 88.52    & 1.47     & 2.58     & 0.94 & 88.80      & 1.71       & 1.94       & 20 & 0.65& 0.22& 9  & 3.53&   10.18    \\
120  & 24  & 3.02&   71 & 2.48  & 0.16(X)& 29 & 2.93  & 0.03(X)& 0.45& 0.59  & 88.54    & 1.48     & 2.55     & 0.94 & 88.81      & 1.69       & 1.96       & 18 & 0.61& 0.22& 8  & 3.49&   15.26    \\
130  & 23  & 2.97&   78 & 2.36  & 0.18(X)& 27 & 2.89  & 0.02(X)& 0.53& 0.59  & 88.55    & 1.49     & 2.52     & 0.94 & 88.81      & 1.68       & 1.97       & 17 & 0.59& 0.22& 8  & 3.44&   18.37    \\
140  & 23  & 2.92&   84 & 2.25  & 0.19(X)& 26 & 2.85  & 0.02(X)& 0.59& 0.59  & 88.57    & 1.50     & 2.49     & 0.94 & 88.81      & 1.68       & 1.98       & 17 & 0.58& 0.22& 8  & 3.40&   20.77    \\
150  & 22  & 2.88&   91 & 2.15  & 0.21(X)& 25 & 2.81  & 0.02(X)& 0.66& 0.59  & 88.57    & 1.51     & 2.47     & 0.94 & 88.81      & 1.67       & 1.98       & 16 & 0.57& 0.22& 8  & 3.35&   23.36    \\
\hline
\multicolumn{24}{c}{$Z = 0.02$}\\
\hline
50   & 33  & 4.19&   25 & 4.29  & 0.72(Y)& 20 & 4.38  & 0.47(Y)& 0.09& 0.58  & 120.87   & 1.71     & 10.59    & 0.92 & 127.37     & 1.80       & 3.04       & 17 & 0.29& 0.07& 12 & 4.62&   2.07     \\ 
60   & 40  & 3.80&   32 & 3.89  & 0.71(Y)& 25 & 3.97  & 0.47(Y)& 0.08& 0.59  & 122.16   & 1.83     & 8.96     & 0.88 & 127.24     & 1.79       & 3.20       & 21 & 0.32& 0.05& 14 & 4.19&   2.04     \\
70   & 39  & 3.55&   39 & 3.56  & 0.98(Y)& 28 & 3.66  & 0.63(Y)& 0.10& 0.58  & 124.28   & 1.89     & 6.46     & 0.91 & 127.31     & 1.85       & 3.05       & 24 & 0.47& 0.05& 13 & 3.94&   2.80     \\
80   & 30  & 3.36&   44 & 3.13  & 0.07(X)& 29 & 3.36  & 0.98(Y)& 0.23& 0.58  & 124.56   & 1.94     & 6.08     & 0.93 & 126.69     & 2.36       & 3.07       & 23 & 0.73& 0.08& 8  & 3.83&   6.96     \\
90   & 21  & 3.27&   50 & 2.80  & 0.13(X)& 26 & 3.17  & 0.03(X)& 0.37& 0.59  & 124.75   & 1.97     & 5.81     & 0.93 & 126.74     & 2.33       & 3.06       & 16 & 0.61& 0.24& 7  & 3.77&   11.57    \\
100  & 20  & 3.19&   56 & 2.59  & 0.16(X)& 22 & 3.12  & 0.02(X)& 0.54& 0.58  & 124.74   & 2.00     & 5.79     & 0.93 & 126.77     & 2.30       & 3.07       & 14 & 0.57& 0.22& 7  & 3.71&   17.18    \\
110  & 19  & 3.14&   62 & 2.40  & 0.19(X)& 20 & 3.09  & 0.01(X)& 0.68& 0.58  & 124.80   & 2.01     & 5.71     & 0.93 & 126.78     & 2.28       & 3.08       & 13 & 0.54& 0.22& 7  & 3.66&   22.12    \\
120  & 18  & 3.08&   68 & 2.27  & 0.21(X)& 19 & 3.04  & 0.01(X)& 0.77& 0.58  & 124.90   & 2.03     & 5.56     & 0.93 & 126.78     & 2.27       & 3.09       & 13 & 0.53& 0.22& 7  & 3.62&   25.40    \\
130  & 18  & 3.04&   74 & 2.15  & 0.23(X)& 18 & 3.01  & 0.01(X)& 0.86& 0.58  & 124.94   & 2.05     & 5.50     & 0.93 & 126.78     & 2.27       & 3.09       & 12 & 0.52& 0.22& 7  & 3.58&   28.59    \\
140  & 17  & 3.00&   80 & 2.04  & 0.24(X)& 18 & 2.98  & 0.01(X)& 0.94& 0.58  & 125.02   & 2.06     & 5.39     & 0.93 & 126.79     & 2.26       & 3.10       & 12 & 0.50& 0.22& 7  & 3.54&   31.54    \\
150  & 17  & 2.95&   85 & 1.98  & 0.25(X)& 18 & 2.93  & 0.01(X)& 0.95& 0.58  & 125.04   & 2.07     & 5.35     & 0.93 & 126.79     & 2.26       & 3.10       & 12 & 0.51& 0.22& 7  & 3.49&   32.34    \\
\hline
\multicolumn{24}{c}{$Z = 0.03$}\\
\hline
50   & 31  & 3.97&   25 & 4.05  & 0.75(Y)& 20 & 4.12  & 0.58(Y)& 0.06& 0.57  & 173.71   & 2.41     & 24.79    & 0.89 & 189.13     & 2.63       & 6.86       & 16 & 0.41& 0.06& 10 & 4.42&   1.54     \\
60   & 34  & 3.64&   31 & 3.68  & 0.89(Y)& 24 & 3.76  & 0.61(Y)& 0.09& 0.59  & 177.86   & 2.59     & 19.80    & 0.89 & 189.36     & 2.72       & 6.48       & 20 & 0.46& 0.05& 10 & 4.07&   2.30     \\
70   & 22  & 3.45&   37 & 3.14  & 0.09(X)& 24 & 3.42  & 0.01(X)& 0.29& 0.57  & 178.02   & 2.66     & 19.53    & 0.92 & 188.90     & 3.34       & 6.19       & 16 & 0.70& 0.13& 6  & 3.99&   8.34     \\
80   & 16  & 3.37&   43 & 2.73  & 0.15(X)& 18 & 3.32  & 0.01(X)& 0.59& 0.57  & 178.43   & 2.71     & 18.98    & 0.92 & 189.07     & 3.27       & 6.08       & 11 & 0.59& 0.21& 5  & 3.98&   17.71    \\
90   & 15  & 3.31&   48 & 2.46  & 0.20(X)& 15 & 3.31  & 0.97(Y)& 0.85& 0.58  & 178.83   & 2.75     & 18.48    & 0.92 & 189.14     & 3.22       & 6.07       & 10 & 0.55& 0.20& 5  & 3.94&   25.78    \\
100  & 14  & 3.26&   54 & 2.25  & 0.23(X)& 13 & 3.27  & 0.97(Y)& 1.03& 0.58  & 179.07   & 2.78     & 18.16    & 0.92 & 189.17     & 3.19       & 6.07       & 9  & 0.52& 0.21& 5  & 3.90&   31.36    \\
110  & 14  & 3.21&   59 & 2.10  & 0.25(X)& 13 & 3.23  & 0.96(Y)& 1.13& 0.58  & 179.21   & 2.81     & 17.97    & 0.92 & 189.20     & 3.18       & 6.05       & 9  & 0.50& 0.21& 5  & 3.86&   34.98    \\
120  & 14  & 3.16&   64 & 1.98  & 0.26(X)& 13 & 3.19  & 0.96(Y)& 1.21& 0.58  & 179.34   & 2.83     & 17.79    & 0.92 & 189.23     & 3.17       & 6.03       & 9  & 0.50& 0.22& 5  & 3.82&   38.01    \\
130  & 14  & 3.13&   70 & 1.86  & 0.28(X)& 13 & 3.15  & 0.96(Y)& 1.29& 0.58  & 179.85   & 2.86     & 17.17    & 0.92 & 189.26     & 3.15       & 6.02       & 8  & 0.49& 0.22& 5  & 3.79&   40.92    \\
140  & 14  & 3.09&   75 & 1.79  & 0.28(X)& 13 & 3.11  & 0.96(Y)& 1.32& 0.58  & 180.13   & 2.88     & 16.82    & 0.92 & 189.28     & 3.15       & 5.99       & 8  & 0.49& 0.22& 5  & 3.74&   42.43    \\
150  & 13  & 3.07&   81 & 1.69  & 0.30(X)& 12 & 3.09  & 0.95(Y)& 1.40& 0.57  & 179.66   & 2.88     & 17.35    & 0.92 & 189.31     & 3.14       & 5.99       & 8  & 0.48& 0.22& 5  & 3.73&   45.21    \\
\hline
\multicolumn{24}{c}{$Z = 0.04$}\\
\hline
50   & 31  & 3.76&   24 & 3.86  & 0.72(Y)& 19 & 3.91  & 0.59(Y)& 0.05& 0.56  & 216.78   & 3.00     & 50.28    & 0.91 & 248.90     & 3.58       & 12.79      & 16 & 0.45& 0.05& 9  & 4.23&   1.25     \\
60   & 26  & 3.48&   31 & 3.33  & 0.04(X)& 21 & 3.52  & 0.88(Y)& 0.19& 0.57  & 222.47   & 3.20     & 43.51    & 0.91 & 248.99     & 3.53       & 12.76      & 17 & 0.70& 0.06& 6  & 4.03&   5.43     \\
70   & 15  & 3.40&   36 & 2.77  & 0.14(X)& 16 & 3.36  & 0.01(X)& 0.59& 0.57  & 224.24   & 3.31     & 41.34    & 0.91 & 248.49     & 4.17       & 12.48      & 9  & 0.62& 0.20& 4  & 4.08&   17.64    \\
80   & 13  & 3.33&   42 & 2.41  & 0.20(X)& 12 & 3.35  & 0.95(Y)& 0.93& 0.56  & 223.61   & 3.34     & 42.02    & 0.91 & 248.74     & 4.08       & 12.31      & 8  & 0.55& 0.20& 4  & 4.04&   27.90    \\
90   & 12  & 3.29&   48 & 2.15  & 0.23(X)& 11 & 3.32  & 0.94(Y)& 1.17& 0.56  & 223.81   & 3.38     & 41.74    & 0.91 & 248.96     & 3.93       & 12.27      & 7  & 0.51& 0.22& 4  & 4.02&   35.32    \\
100  & 12  & 3.24&   53 & 1.98  & 0.25(X)& 11 & 3.28  & 0.93(Y)& 1.30& 0.56  & 224.36   & 3.42     & 41.06    & 0.91 & 249.60     & 3.43       & 12.19      & 7  & 0.49& 0.22& 4  & 3.97&   39.55    \\
110  & 12  & 3.20&   57 & 1.86  & 0.27(X)& 10 & 3.24  & 0.92(Y)& 1.39& 0.56  & 224.93   & 3.45     & 40.36    & 0.91 & 250.00     & 3.14       & 12.13      & 7  & 0.48& 0.23& 4  & 3.94&   42.72    \\
120  & 12  & 3.17&   63 & 1.74  & 0.29(X)& 10 & 3.21  & 0.91(Y)& 1.48& 0.56  & 225.41   & 3.48     & 39.78    & 0.91 & 250.02     & 3.16       & 12.08      & 7  & 0.48& 0.23& 4  & 3.91&   45.93    \\
130  & 12  & 3.13&   68 & 1.64  & 0.30(X)& 10 & 3.18  & 0.91(Y)& 1.55& 0.56  & 225.65   & 3.51     & 39.47    & 0.91 & 250.06     & 3.16       & 12.02      & 7  & 0.47& 0.23& 4  & 3.87&   48.60    \\
140  & 12  & 3.11&   74 & 1.54  & 0.31(X)& 10 & 3.16  & 0.90(Y)& 1.62& 0.56  & 225.38   & 3.52     & 39.76    & 0.91 & 250.11     & 3.15       & 11.98      & 7  & 0.47& 0.23& 4  & 3.85&   51.32    \\
150  & 12  & 3.08&   78 & 1.48  & 0.32(X)& 10 & 3.14  & 0.90(Y)& 1.66& 0.56  & 226.07   & 3.55     & 38.93    & 0.91 & 250.16     & 3.15       & 11.93      & 7  & 0.47& 0.23& 4  & 3.83&   52.91    \\
\enddata
\end{deluxetable*} 
\end{longrotatetable}


\bibliography{sample631}{}
\bibliographystyle{aasjournal}



\end{document}